\documentclass[11pt]{article}

\usepackage[T1]{fontenc}
\usepackage[utf8]{inputenc}

\global\arraycolsep=2pt

\usepackage[a4paper,top=30.6mm,bottom=38.6mm,left=34.6mm,right=34.6mm,footskip=1.3cm]{geometry}

\usepackage[nofancyfonts]{packages}
\allowdisplaybreaks

\acrodef{ybd}[\textsc{yb} deformation]{Yang--Baxter deformation}
\acrodef{nc}[\textsc{nc}]{non-commutativity}
\acrodef{sym}[\textsc{sym}]{super Yang--Mills}
\acrodef{himr}[\textsc{himr}  background]{Hashimoto--Itzhaki--Maldacena--Russo background~\cite{hep-th/9907166,hep-th/9908134}}
\acrodef{cybe}[\textsc{cybe}]{classical Yang--Baxter equation}

\DeclareMathOperator{\projop}{\Pi}
\newcommand*{\proj}[1]{\ensuremath{\projop^{\text{#1}}}}

\newcommand*{\setR}{\ensuremath{\mathbb{R}}}
\newcommand*{\setZ}{\ensuremath{\mathbb{Z}}}
\newcommand*{\del}{\mathop{\mathrm{{}\partial}}\mathopen{}}
\newcommand*{\tIIA}{type~\textsc{iia}\xspace}
\newcommand*{\tIIB}{type~\textsc{iib}\xspace}

\usepackage{fancyhdr}
\newcommand{\preprint}[1]{%
  \fancypagestyle{empty}{%
    \fancyhf{} %
    \fancyheadoffset[R]{\marginparsep+\marginparwidth}
    \fancyhead[R]{{#1}}
    \renewcommand{\headrulewidth}{0pt}
    \renewcommand{\footrulewidth}{0pt}
  }
}

\newcommand*{\AdS}[1]{\ensuremath{\mathrm{AdS}_{#1}}}

\begin{document}

\numberwithin{equation}{section}

\preprint{KUNS-2721}

\begin{center}
{\Huge \textbf{Killing spinors from classical $r$--matrices} }
\vspace*{1.5cm}\\
{\large 
Domenico Orlando$^{\sharp,\flat}$, 
Susanne Reffert$^{\flat}$, 
\\ 
Yuta Sekiguchi$^{\flat}$
and Kentaroh Yoshida$^{\ast}$
} 
\end{center}
\vspace*{0.2cm}
\begin{center}
  $^{\sharp}${\it INFN, sezione di Torino and Arnold--Regge Center\\
  via Pietro Giuria 1, 10125 Torino, Italy}
\vspace*{0.25cm}\\ 
$^{\flat}${\it Institute for Theoretical Physics, 
Albert Einstein Center for Fundamental Physics, 
University of Bern, Sidlerstrasse 5, CH-3012 Bern, 
Switzerland 
}
\vspace*{0.25cm}\\ 
$^{\ast}${\it Department of Physics, Kyoto University, \\ 
Kyoto 606-8502, Japan} 
\end{center}
\vspace{0.3cm}

\begin{abstract}
The \ac{ybd} is a systematic way of performing integrable deformations 
of two-dimensional symmetric non-linear sigma models. The deformations 
can be labeled by classical $r$-matrices satisfying the classical \textsc{yb} equation. 
This \ac{ybd} is also applicable to type IIB superstring theory defined on AdS$_5\times S^5$. 
In this case, a simple class of \acp{ybd} is associated with TsT transformations of 
string backgrounds. The data of this transformation is encoded 
in an antisymmetric bi-vector $\Theta$ (which is often called $\beta$ field or non-commutativity parameter). 
In this article, we give a simple recipe for obtaining the explicit expression for the Killing spinors 
of the TsT transformed background starting from $\Theta$. 
We moreover discuss the M-theory 
equivalent of the TsT transformation, allowing us to also give Killing spinors of 11-dimensional 
backgrounds. We discuss examples of TsT transformed backgrounds starting from flat space, 
$\AdS{5}\times S^5$ and $\AdS{7}\times S^4$. We find that in this way 
we can relate the $\Omega$-deformation to \acp{ybd}.
\end{abstract}

\setcounter{footnote}{0}
\setcounter{page}{0}
\thispagestyle{empty}

\newpage
\tableofcontents
\newpage
\section{Introduction} 
\label{sec:introduction}

An intriguing property of the AdS/CFT correspondence~\cite{hep-th/9711200} 
is the integrable structure behind it. It is so powerful that the conjectured duality 
can be tested even in the non-BPS region by employing integrability techniques 
(for a comprehensive review, see~\cite{arXiv:1012.3982}).  
This integrable structure is encoded in the target space geometry 
on the string-theory side. For a typical example of AdS/CFT, 
type IIB string theory on AdS$_5\times S^5$ is considered, 
and then the target spacetime is represented by a symmetric coset~\cite{hep-th/9805028} 
of the form
\begin{eqnarray}
\frac{PSU(2,2|4)}{SO(1,4) \times SO(5)}\,. 
\end{eqnarray}
This supercoset enjoys a $\mathbb{Z}_4$-grading property~\cite{hep-th/0305116} 
and it ensures the classical integrability of the string theory in the sense of 
kinematical integrability, which means the existence of classical Lax pair. 

A significant research direction is considering integrable deformations of AdS/CFT 
and constructing a variety of examples to which powerful integrability techniques 
can be applied. To push forward in this direction, it is desirable to follow a systematic approach 
as the case-by-case study is tedious and the {\it universal} structure cannot be discovered. 
It is by now well-recognized that a systematic method of performing integrable deformations called 
\acf{ybd}~\cite{hep-th/0210095,arXiv:0802.3518,arXiv:1308.3581,arXiv:1501.03665} is well-suited to studying integrable deformations 
of AdS/CFT~\cite{arXiv:1309.5850,arXiv:1402.6147}. In the \ac{ybd} approach, 
the deformations can be labeled by classical $r$-matrices satisfying 
the \ac{cybe}. In particular, 
by performing the supercoset construction~\cite{Kyono:2016jqy,Arutyunov:2015qva},
the associated deformed string backgrounds can be derived.\footnote{For the Weyl invariance 
of the deformed background, see~\cite{Arutyunov:2015mqj, Sakamoto:2017wor}.}

In this article, we will concentrate on \acp{ybd} based on the homogeneous 
\ac{cybe}~\cite{arXiv:1402.6147, arXiv:1501.03665}. In this case, there exists 
an important condition called {\it the unimodularity condition} 
for the classical $r$-matrices~\cite{Borsato:2016ose}.\footnote{The authors of~\cite{Borsato:2016ose} also compute the number of supersymmetries preserved by \ac{ybd}. Their results are in agreement with our discussion in Section~\ref{sec:Killing-spinors}.}
When this condition is satisfied, the associated deformed background satisfies the on-shell condition 
of the usual supergravity. If the condition is not satisfied, the resulting background cannot be solutions of 
the usual supergravity, but of a generalized supergravity~\cite{Arutyunov:2015mqj}.
In other words,  the unimodularity condition ensures the on-shell condition of 
standard supergravity. For many examples of classical $r$-matrices, 
the complete solutions of generalized supergravity have been obtained  
in~\cite{Orlando:2016qqu, arXiv:1710.06849}. 

It should be noted here that the appearance of generalized supergravity is not surprising. 
The $\kappa$-symmetry constraints in the Green--Schwarz formulation of superstring theories 
lead to the equations of motion of generalized supergravity~\cite{Wulff:2016tju}. 
Hence, although generalized supergravity was artificially introduced in order to support 
an $\eta$-deformed AdS$_5\times S^5$ background as a solution~\cite{Arutyunov:2015mqj}, 
its fundamental origin has in the meantime been uncovered. 

In the following, for the sake of simplicity, we will focus on the unimodular case, in which   
there is an interesting connection between classical $r$-matrices and 
TsT transformations, where a TsT transformation consists of a T-duality, 
a shift of the dualized coordinate and another T-duality back. 
This relation has been discovered in a series of papers~\cite{arXiv:1404.1838, 
arXiv:1404.3657,arXiv:1412.3658, arXiv:1502.00740}\footnote{For \acp{ybd} 
of Minkowski spacetime, see~\cite{Matsumoto:2015ypa, arXiv:1510.03083, 
arXiv:1512.00208,arXiv:1710.06849}.}. {All of the Abelian} classical $r$-matrices
are associated with TsT transformations~\cite{arXiv:1608.08504}.

As another interesting property, classical $r$-matrices utilized for \acp{ybd} of AdS$_5$ are
closely related to non-commutativity on the dual gauge-theory side as originally pointed out 
in~\cite{arXiv:1404.3657}. This observation has been further developed in subsequent works~\cite{arXiv:1504.05516, arXiv:1506.01023, arXiv:1610.05677,  arXiv:1702.02861,  arXiv:1705.02063, arXiv:1705.07116, arXiv:1708.03163, arXiv:1710.06849,  arXiv:1803.05903, Hoare:2016wca, Hassler:2017yza, Lust:2018jsx, Bakhmatov:2018apn}.
In summary, the non-commutativity is schematically related to the classical $r$-matrix via
\begin{equation}
\Theta^{mn} = r^{mn} ,
\end{equation}
where $\Theta^{mn}$ is an antisymmetric bi-vector\footnote{In {this paragraph, 
we call} $\Theta^{mn}$ non-commutativity for brevity.} obtained from the non-commutativity parameter by taking a certain limit. 
When the non-commutativity parameter
is constant, this relation is obvious. As shown in~\cite{arXiv:1404.3657}, the Yang--Baxter deformed background is nothing but the \ac{himr} and the Moyal plane discussed by Seiberg and Witten~\cite{hep-th/9908142} is associated 
(i.e., $\Theta^{mn}$ is just a constant).\footnote{If $\Theta$ is not 
a constant, the argument should be involved as discussed in~\cite{hep-th/9908134}.}
Then the unimodularity condition corresponds to the divergence-less condition for $\Theta^{mn}$~\cite{arXiv:1702.02861,arXiv:1705.02063},
\begin{equation}
\nabla_m \Theta^{mn} = 0. 
\end{equation}
For the non-unimodular case, an extra, non-dynamical vector appears 
on the right-hand side as a source term~\cite{arXiv:1702.02861,arXiv:1705.02063}.

The main subject of this article is to revisit \acp{ybd} of \tIIB string theory  
on AdS$_5\times S^5$ with unimodular classical $r$-matrices from the viewpoint 
of the Killing spinor. In particular, we clarify the concise recipe for reconstructing Killing spinors for \textsc{yb}
deformed backgrounds from the associated classical $r$-matrix. After giving the recipe, 
we are in the position to identify the classical $r$-matrix from the already known Killing spinor. 
Along this direction, we identify classical $r$-matrices associated to $\Omega$-deformations.

Before going into details, we outline our strategy here. 
Let us first recall what a TsT transformation is. 
Suppose that \tIIB string theory is compactified on $T^2$. This system has 
an $SL(2,\setZ) \times SL(2,\setZ)$ symmetry. The first $SL(2,\setZ)$ 
acts geometrically  on the $T^2$ while the second one is non-geometric, 
acting on the K\"ahler parameter $\tau =B_{12} + i \sqrt{g_{(12)}}$, 
where $\alpha_1, \alpha_2$ are the directions of the torus. 
The TsT transformation realizes the transformation $\tau \to \tau/(1+\lambda\tau)$, 
where $\lambda$ is a real shift parameter. 
Because $\lambda$ is real, this transformation is not a symmetry. 
The resulting background is therefore inequivalent to the initial one.

A similar construction can also be realized in M-theory compactified on $T^3$, 
making use of the $SL(3,\setZ) \times SL(2,\setZ)$ symmetry, 
where again the $SL(3,\setZ)$ acts geometrically on $T^3$, 
while the $SL(2,\setZ)$ acts on $\tau^M =C_{123} + i \sqrt{g_{(123)}}$, 
with $\alpha_1, \alpha_2, \alpha_3$ being again the torus directions.

For a Killing spinor to be compatible with the TsT transformation, 
it must not depend on any of the torus directions.  
Our main result is a simple recipe for translating the {bi-vector $\Theta^{mn}$}
encoding the transformation data into an explicit formula for the Killing spinors of the background resulting from the TsT transformation:\footnote{The exponential factor below is $\Omega$ introduced in \cite{arXiv:1803.05903}. 
We appreciate Yuho Sakatani for this point.}
\begin{align}
  \epsilon_{L}^{\text{(fin)}} &= \proj{TsT} \epsilon_L^{\text{(in)}} , &
  \epsilon_{R}^{\text{(fin)}} &=  e^{\omega(\Theta) \, \Theta^{\mu\nu}\hat{\Gamma}_{\mu}\hat{\Gamma}_{\nu}}\proj{TsT} \epsilon^{\text{(in)}}_{R}\,,
\end{align}
where \(\epsilon^{\text{(in)}}\) are the Killing spinors in the initial background, \(\omega(\Theta)\) is a normalization factor and \(\proj{TsT}\) is an appropriate projector that is completely determined by \(\Theta\).

We find that the $\Omega$-deformation of flat space, realized via the fluxtrap construction~
\cite{arXiv:1106.0279, Reffert:2011dp, arXiv:1309.7350}, is a particular case of this class of 
\acp{ybd}. The relationship between the $\Omega$-deformation 
and non-commutativity has already been observed in the past~\cite{Aganagic:2003qj, 
Dijkgraaf:2008ua, Mironov:2009dv, arXiv:1204.4192}.
Here we observe that it can be understood in terms of an $SL(2, \setR)$ transformation, 
relating it on the one hand to integrability and on the other hand to non-commutativity.

This paper is organized as follows. In Section~\ref{sec:R-matrix}, we review the TsT transformation, 
the associated classical $r$-matrix and bi-vector $\Theta$. In Section~\ref{sec:Killing-spinors}, 
we describe generalities for the Killing spinors in type II supergravity. 
In particular, we provide a recipe for obtaining the explicit expression for the Killing spinors 
preserved by different types of TsT transformations both for flat space (Sec.~\ref{sec:flat}) 
and $\AdS{5}\times S^{5}$ (Sec.~\ref{sec:AdS5}). 
In Section~\ref{sec:examples} we investigate various examples of \textsc{yb}-deformed backgrounds 
characterized by classical $r$-matrices, {and then discuss} the $\Omega$-deformation of flat space.
In Section~\ref{sec:conclusions}
we give some concluding remarks and an outlook. 
In the appendices, the conventions and computational details for the Killing spinors are presented.

\section{Set-up of terminologies and techniques}
\label{sec:R-matrix}

In this section, we introduce the TsT transformation and the associated anti-symmetric bi-vector $\Theta$ 
via a field redefinition. These are closely related to classical $r$-matrix employed in the \ac{ybd}.  

In Section~\ref{sec:TsT-NC}, we introduce what a TsT transformation is and describe 
the relation between a TsT transformation and an anti-symmetric bi-vector $\Theta$. 
Then, through a field redefinition, this $\Theta$ is associated with a classical $r$-matrix 
which is employed in the \ac{ybd}. 
Section~\ref{NC:sec} provides a list of non-commutative spaces for later convenience. 
This is relevant to deformations of flat space and AdS space. 
Finally, Section~\ref{sec:TsT-M} briefly explains how one can see TsT transformation in the context of M-theory. 
This is also related to our later discussion. 

\subsection{TsT, $\Theta$ and classical $r$-matrix} \label{sec:TsT-NC}

Here, let us introduce the TsT transformation, the associated anti-symmetric bi-vector $\Theta$ and the classical $r$-matrix. 

\subsubsection*{TsT transformation} 

First of all, we will introduce the definition of the TsT transformation. 

A TsT-transformation consists of a combination of two T-dualities and a shift transformation. 
Suppose that there exist two isometry directions $\alpha_1$ and $\alpha_2$ 
in the background of our interest. Then we define the TsT transformation as
\begin{equation}
(\alpha_1, \alpha_2)_{\lambda} \equiv
    \begin{cases}
    \text{T-duality \(\alpha_1 \to \tilde \alpha_1\)} \\
    \text{shift \(\alpha_2 \to  \alpha_2 + \lambda \tilde \alpha_1 \)} \\
    \text{T-duality on \(\tilde \alpha_1\)} .
  \end{cases}
\end{equation}
The transformations $(\alpha_1, \alpha_2)_{\lambda}$ and $(\alpha_2, \alpha_1)_{\lambda}$ must give the same result because there is another, geometric, commuting $SL(2,\setR)$ that permutes the two directions. 

A well-known example is the Lunin--Maldacena 
background~\cite{hep-th/0502086}, which is dual to $\beta$-deformed supersymmetric 
Yang--Mills theory~\cite{Leigh:1995ep}. This is a TsT transformation of $S^5$ 
and can also be interpreted as a \ac{ybd}~\cite{arXiv:1404.1838}.  
Note here that all of the \acp{ybd} of $S^5$ are labeled by unimodular 
classical $r$-matrices~\cite{Borsato:2016ose}\footnote{Deformations of $S^5$ correspond to 
deformations of the superpotential for the scalar fields in the $\mathcal{N}=4$ \ac{sym}. Nevertheless, this case 
is often called ``non-commutative'' since the superpotential receives a $\star$-product-like deformation~\cite{hep-th/0502086}. }. 

Another example is the \ac{himr}.  This background can also be seen as a \ac{ybd}. 
Thus, as originally observed in~\cite{arXiv:1404.3657}, there is an intimate connection among TsT transformations, 
non-commutativity, and classical $r$-matrices in the \ac{ybd}.

\subsubsection*{An antisymmetric bi-vector $\Theta$}

Next, we will introduce the anti-symmetric bi-vector $\Theta$ for later purposes. This quantity plays a central role 
in our analysis. 

This bi-vector $\Theta$ can be introduced by performing the following field redefinition 
 (for the NS-NS sector)~\cite{Duff:1989tf,hep-th/9908142}: 
\begin{eqnarray}
G_{mn} &=& (g - B g^{-1}B)_{mn}, \\ 
\Theta^{mn} &=& -((g+B)^{-1}B(g-B)^{-1} )^{mn}, \\ 
G_s &=& g_s \left(\frac{\det(g+B)}{\det g}\right)^{1/2}. 
\end{eqnarray}
Here $g_{mn}$, $B_{mn}$ and $g_s$ are the closed string metric, NS-NS two-form and closed string coupling, respectively. 
$G_{mn}$ and $G_s$ are the open string metric and coupling, respectively, and $\Theta$ is an anti-symmetric bi-vector. 

It should be remarked that this bi-vector $\Theta$ is called a $\beta$ field~\cite{Duff:1989tf} 
or non-commutativity~\cite{hep-th/9908142}. For the former, the $\beta$ field was introduced in the context of a duality rotation. This quantity is closely related to the non-geometric Q-flux~\cite{Grana:2008yw, Andriot:2011uh} and 
\acp{ybd} can capture non-geometric aspects of the geometry like T-folds~\cite{arXiv:1710.06849}.  
But for the latter terminology we must be careful. 
For some special cases like deformations of flat space and AdS space, this $\Theta$ may be understood 
as a non-commutativity of the associated non-commutative plane (supposing a holographic relation 
in the case of deformations of AdS). But in general, this terminology may cause confusion. In the following, 
we will call $\Theta$ simply a bi-vector. In case that a relation between $\Theta$ and non-commutativity 
is anticipated, we will say that $\Theta$ measures a non-commutativity. 

Even if an interpretation of $\Theta$ as a non-commutativity is expected, we must still be careful for 
the definition of terminology. In the case of deformations of AdS, there exist two kinds of non-commutativity: 
1) the bulk non-commutativity\footnote{Note that this word does not mean the non-commutativity of 
the bulk spacetime coordinates, but implies the bulk counter-part of the non-commutativity on the boundary.} 
and 2) the boundary non-commutativity. The former is nothing but $\Theta$ 
and the latter is a non-commutativity of the space on which the dual gauge theory lives.

\paragraph{Relation to YB deformations} 

The field redefinition itself is quite general and does not depend on the \ac{ybd}. 
However, if we perform this field redefinition for 
\textsc{yb}-deformed backgrounds with the homogeneous \ac{cybe}, then  
one can see a very interesting result~\cite{arXiv:1506.01023, arXiv:1610.05677,  arXiv:1702.02861,  arXiv:1705.02063, arXiv:1705.07116}: 1) the open string metric $G_{mn}$ becomes the original undeformed metric, 
2) the open string coupling becomes constant and 3) all of the information of the deformation is encoded 
into the bi-vector $\Theta^{mn}$. 

That is why the bi-vector $\Theta^{mn}$ plays a significant role in the following discussion.  

\subsubsection*{Classical $r$-matrix from $\Theta$}

In fact, the bi-vector $\Theta$ is related to the isometry directions $\alpha_1$ and $\alpha_2$ associated to the transformation $(\alpha_1, \alpha_2)_{\lambda}$. It is given by (up to signs and numerical factors) 
\cite{arXiv:1506.01023, arXiv:1610.05677,  arXiv:1702.02861,  arXiv:1705.02063, arXiv:1705.07116}
\begin{equation}
\begin{aligned}
\Theta = \frac{1}{2}\Theta^{\mu\nu}\partial_{\mu} \wedge \partial_{\nu} = \lambda \partial_{\alpha_1} \wedge \partial_{\alpha_2} .
\end{aligned}
\end{equation}
The last expression is nothing but a classical $r$-matrix expressed 
in terms of Killing vector in the bulk. 

In the case of the \ac{himr}
with a non-commutativity for the $x^1$ and $x^2$-directions, only $\Theta^{12}$ is a non-vanishing constant, 
and the other components are zero. Thus the associated classical $r$-matrix is given by~\cite{arXiv:1404.3657}
\begin{equation}
\begin{aligned}
\Theta = \Theta^{12}\partial_{1} \wedge \partial_{2} 
= \lambda \partial_{1} \wedge \partial_{2}, 
\end{aligned}
\end{equation}
where $\partial_1$ and $\partial_2$ are Killing vectors for the bulk background. 

Recall that the bi-vector $\Theta$ is the bulk non-commutativity 
and different from the non-commutativity of the dual gauge theory. 
But in the case that $\Theta^{mn}$ is constant, the two non-commutativities 
coincide with each other. 

In the following discussion, we will encounter more general, non-constant non-commutativities. 
Hence in the next subsection, we will summarize possible non-commutative spaces relevant to our 
later discussion.

\subsection{Non-commutative spaces\label{NC:sec}}

It would be helpful to summarize possible non-commutative spaces relevant to our later discussion.  
In the following, we will follow the work~\cite{Wess:2003da} (see also~\cite{Lukierski:2005fc}). 

The non-commutativity of the spacetime variables is formulated based on the commutation relations 
\begin{equation}
[\hat{x}^{\mu}, \hat{x}^{\nu}] = \frac{1}{\kappa^2} \theta^{\mu\nu}(\kappa\hat{x})\,,
\end{equation}
where $\kappa$ is a mass-like parameter. Here $\hat{x}^{mu}$ should be interpreted as an operator 
and hence $\theta$ is a function of operators. This function is defined formally as a Taylor series expansion like 
\begin{eqnarray}
\theta(\hat{x}) = \theta^{\mu\nu}_{(0)} + \kappa \theta^{\mu\nu}_{(1)~\rho} \hat{x}^{\rho} 
+ \kappa^2 \theta^{\mu\nu}_{(2)~\rho\sigma}\hat{x}^{\rho}\hat{x}^{\sigma} + \cdots. 
\label{expansion}
\end{eqnarray} 
The non-commutativity $\theta^{\mu\nu}(\hat{x})$ is constrained by various conditions, 
such as the Jacobi identity and antisymmetry.

Let us list up below examples of non-commutative spaces relevant to our later discussion. 
These are realized as special cases in the expansion (\ref{expansion}). 
\begin{enumerate}
  \renewcommand{\labelenumi}{(\roman{enumi})}
\item The canonical case:\\
The canonical relation is the same as in quantum mechanics, 
but valid within the coordinate space.
\begin{equation}
\comm{\hat{x}^{\mu}}{\hat{x}^{\nu}} = i \theta_{(0)}^{\mu\nu}\,,
\end{equation}
Note that this non-commutative structure emerged 
in the context of string theory first in~\cite{Chu:1998qz} and was elaborated in~\cite{hep-th/9908142}.

\item The Lie algebra case:\\
The \ac{nc} parameter is supplied with one more leg, and regarded 
as the standard structure constant $\Theta^{\mu\nu}{}_{\rho}$ as defined in the Lie algebra~\cite{Jurco:2000ja, LUKIERSKI1991331}. 
Then the commutator yields on the right-hand-side a linear term in $\hat{x}$\,,
\begin{equation}
\comm{\hat{x}^{\mu}}{\hat{x}^{\nu}} = i \theta^{\mu\nu}_{(1)~\rho}\hat{x}^{\rho}\,.
\end{equation}

\item The quantum spaces:\\
The \ac{nc} parameter is uplifted to include one more index, so that one finds a quadratic term in the space variable:
\begin{equation}
\comm{\hat{x}^{\mu}}{\hat{x}^{\nu}} = i \theta\indices{^{\mu\nu}_{(2)~\rho\sigma}}\hat{x}^{\rho}\hat{x}^{\sigma}\,.
\end{equation}
\end{enumerate}

We assume that these non-commutative spaces are realized on the boundary when considering deformations of 
AdS. The bulk counter-part $\Theta$ is different from $\theta$ in general. However, the bulk non-commutativity $\Theta$ 
is realized as the leading term in the expansion of $\theta$ in terms of the deformation parameter (different from $\kappa$) 
as shown in~\cite{arXiv:1702.02861, arXiv:1705.02063}.

\subsection{TsT in M-theory}\label{sec:TsT-M}

In~\cite{hep-th/0502086}, the TsT transformation \((\alpha_1, \alpha_2)_\lambda\) was introduced as a string-theoretical way to realize the \(SL(2, \setR)\) transformation \(\tau \to \tau/\pqty{1+ \lambda \tau}\) on the parameter \(\tau = B_{12} + i \sqrt{g_{(12)}}\), where \(g_{(12)}\) is the volume of the two-torus generated by \(\alpha_1\) and \(\alpha_2\).
We present the extension of this construction to M-theory which we will use  in order to capture also higher-dimensional examples such as $\AdS{7} \times S^4$.\footnote{The example of $\AdS{4}\times S^7$ has already appeared in~\cite{hep-th/0502086}.}
We do this in view of eventually extending the study of integrable deformations to an M-theory setting.

Consider M-theory compactified on the three-torus generated by \(\alpha_1\), \(\alpha_2\) and \(\alpha_3\).
The resulting eight-dimensional theory has a discrete \(U\)-duality group \(U = SL(3, \setZ) \times SL(2, \setZ)\).
The low-energy supergravity theory has a non-compact symmetry \(SL(3, \setR) \times SL(2,\setR)\), which --- as usual --- can be used as a group of solution-generating transformations.

From the point of view of eleven-dimensional supergravity, the \(SL(3, \setR)\) is the geometric group of large coordinate transformations on the torus, while the non-geometric \(SL(2, \setR)\) acts as a linear fractional transformation on the parameter
\begin{equation}\label{eq:tauM}
  \tau^{M} = C_{123} + i \sqrt{g_{(123)}},
\end{equation}
where \(g_{(123)}\) is the volume of the three-torus.
In order to realize a generic transformation
\begin{equation}
  \tau^M \to \frac{a + b\tau^M}{c + d  \tau^M}
\end{equation}
in terms of operations on the eleven-dimensional background, we first notice that if we compactify M-theory on \(\alpha_3\), we obtain \tIIA string theory on \(T^2\), which has the same \(SL(2)\) action on \(\tau\), which now is defined as
\begin{equation}
  \tau^M = \tau^{IIA} = B_{12} + i \sqrt{g_{(12)}} .
\end{equation}
Now we can realize a generic \(SL(2, \setR)\) transformation on \(\tau^{IIA}\) via the usual series of dualities and shifts, and finally lift the result to M-theory.

For concreteness, the transformation
\begin{equation}
  \tau^M \to \frac{\tau^M}{ 1 + \lambda \tau^M}
\end{equation}
is realized by the following series of transformations:
\begin{equation}
  (\alpha_1, \alpha_2, \alpha_3)_\lambda =
  \begin{cases}
    \text{reduction on \(\alpha_3\)} \\
    \text{T-duality \(\alpha_1 \to \tilde \alpha_1\)} \\
    \text{shift \(\alpha_2 \to \alpha_2 + \lambda \tilde \alpha_1\)} \\
    \text{T-duality on \(\tilde \alpha_1\)} \\
    \text{lift on \(\tilde \alpha_3\)}.
  \end{cases}
\end{equation}
Since the actions of \(SL(3, \setR)\) and \(SL(2, \setR)\) commute, the final result is independent of the order of the variables \(\alpha_1\), \(\alpha_2\) and \(\alpha_3\).

This construction is analogous to the one discussed in~\cite{Aharony:1996wp,Ganor:1996zk} where the transformation \(\tau^M \to - 1/ \tau^M\) was realized as reduction on \(\alpha_3\), T-duality on \(\alpha_1\) and \(\alpha_2\) and lift on \(\tilde \alpha_3\) (this is the so-called ``M-theory T-duality'').

\section{TsT-transformed Killing spinors}
\label{sec:Killing-spinors}

\subsection{General construction}\label{sec:general}

We have seen how \acp{ybd} can be understood in terms of TsT transformations.
In this section we compute the supersymmetry properties of 
the resulting ten-dimensional backgrounds.

Following the construction in~\cite{arXiv:1106.0279}, we can see 
how the Killing spinors of a given background transform under T-duality.
For concreteness, we start from a \tIIB background with a doublet of 
Majorana--Weyl Killing spinors \((k_L^{(1)}, k_R^{(1)})\).
We pick a local frame and let \(u\) be a compact isometry, 
\emph{i.e.} let  \(\del_u\) be a Killing vector for the metric and suppose that 
the Killing spinors do not depend on \(u\) (in this frame).
Then, after a T-duality in \(u\), the resulting \tIIA background admits the Killing spinors~\cite{Bergshoeff:1994cb}
\begin{equation}
  \begin{cases}
    k_L^{(2)} =  k_L^{(1)}, \\
    k_R^{(2)} =  \Gamma_u k_R^{(1)},
  \end{cases}
\end{equation}
where \(\Gamma_u\) is the gamma matrix in the \(u\) direction normalized to unity\footnote{It was observed in~\cite{arXiv:1106.0279} that this matrix remains invariant under the duality.}
\begin{equation}
  \Gamma_u = \frac{1}{\pqty{g_{uu}}^{1/2}} e\indices{^A_u} \Gamma_A ,
\end{equation}
where \(g\) and \(e\) are, respectively, the metric and the vielbein.
In the following, the superscript $(1)$ refers to quantities before T-duality and the superscript $(2)$ to quantities after T-duality.
A useful form for the vielbein after the T-duality transformation was given in~\cite{Bergshoeff:1994cb}:
\begin{equation}
  \begin{cases}
    {e^{(2)}}\indices{^A_\sigma} = {e^{(1)}}\indices{^A_\sigma} - \frac{{g^{(1)}}_{\sigma u} + {B^{(1)}}_{\sigma u}}{{g^{(1)}}_{uu}} {e^{(1)}}\indices{^A_u} & \text{for \(x^\sigma \neq u\)}, \\
    {e^{(2)}}\indices{^A_u} = {e^{(1)}}\indices{^A_u} .
  \end{cases}
\end{equation}

These simple expressions make it very easy to write the Killing spinors for backgrounds resulting from a TsT transformation \((u,v)_\lambda\).
If the Killing spinors of the initial metric \emph{do not} depend on the variables \(u\) and \(v\), then
\begin{equation}
  \label{eq:Killing-after-TsT}
  \begin{cases}
    k^{\text{(fin)}}_L = k^{\text{(in)}}_L , \\
    k^{\text{(fin)}}_R =  \Gamma^{\text{(fin)}}_u \Gamma^{\text{(in)}}_u k^{\text{(in)}}_R .
  \end{cases}
\end{equation}
As was argued in~\cite{Bergshoeff:1994cb}, if \(k^{\text{(in)}}\) depends explicitly on \(u\) then --- at the level of supergravity --- it is not preserved under T-duality.
The simplest toy example is given by \(\setR^2 \) in the polar coordinate frame \({\dd{s}}^2 = {\dd{r}}^2 + r^2 {\dd{u}}^2\).
The initial Killing spinors have the form \(K = \exp[ u/2 \Gamma_{12}] \eta_0\), where \(\eta_0\) is a constant spinor.
Since all the spinors depend on \(u\), we do not expect them to survive a T-duality in this direction.
In fact, the dilaton background \({\dd{s}}^2 = {\dd{r}}^2 + (1/r^2) {\dd{u}}^2 \), \(\Phi = - \log u\) breaks all supersymmetries.\footnote{The situation is different in string theory where supersymmetry is not broken by T-duality but is realized in terms of extra modes that do not appear in the perturbative supergravity description~\cite{Duff:1997qz, Duff:1998us}.}

If the Killing spinor depends explicitly on \(v\) (\emph{i.e.} the direction in which the shift is performed), we would like to argue that supersymmetry is broken by the boundary conditions.
To see this, observe that if \(k^{\text{(in)}}\) depends on \(v\), it must be periodic,
\begin{equation}
  k(v) = k(v + 2 \pi) .
\end{equation}
Under the shift operation, \(v\) goes to \(v - \lambda \tilde u\) (where \(\tilde u\) is the T-dual variable to \(u\)) and now we have \(k( v - \lambda \tilde u)\) that has to be invariant under both \(v \to v + 2 \pi\) \emph{and} \(\tilde u \to \tilde u + 2 \pi\).
It is easy to see that unless \(\lambda \) is an integer, this transformation is not an invariance and supersymmetry is not compatible with the periodic boundary conditions.
Interestingly, if we ignore this fact and naively apply the formulas in Eq.~\eqref{eq:Killing-after-TsT}, we find that the resulting spinors \emph{do not satisfy} the Killing equations in the final frame.
In other words, the second T-duality turns a \emph{global} periodicity condition into a \emph{local} one. 

In conclusion, only the Killing spinors that do not depend on either direction  are compatible with the global conditions necessary for the deformation.\footnote{This notion of dependence can be made frame-invariant using the concept of Kosmann Lie derivative for spinors~\cite{Kosmann1971,Kelekci:2014ima}. We would like to thank Eoin Colgáin for discussions on this point}

\subsection{Recipe for flat space}\label{sec:flat}

We want to find a systematic way of classifying TsT transformations in an invariant way which gives rise to the same classification of the bi-vector $\Theta$ used in Section~\ref{sec:TsT-NC}, namely the canonical, Lie-algebra and quantum space cases.

\paragraph{Killing spinors in different coordinate frames.}\

Let us consider TsT transformations of flat space as a warm-up example. In the simplest case, space-time is of the type $\setR^{1,7} \times T^2$ and the TsT transformation is performed on the two directions generating the torus. This is the standard Seiberg--Witten case~\cite{hep-th/9908142}. Here, the Killing spinors depend neither on $u$ nor $v$ and no symmetries are broken. 

\bigskip

The next simplest situation is $\setR^{1,8} \times S^1$  and the TsT transformation is $(u,\phi)_\lambda$, where $u$ generates the $S^1$ and $\del_\phi$ is the generator of rotations in a two-dimensional plane $\setR^2 \subset \setR^{1,8}$. The first step is to find an appropriate frame in which $u$ and $\phi$ appear explicitly. The simplest choice is 
\begin{equation}
\label{eq:flat-background}
	\dd{s^2} = -(\dd{x^0})^2 + \sum_{m=1}^{6}(\dd{ x^{m}})^2 + \dd{ \rho^2 } + \rho^2 \dd{ \phi^2} + \dd{ u^2}.
\end{equation}
Then, the Killing spinors have the form 
\begin{equation}
\label{eq:flat-Killing-spinor1}
  K = \exp[ \frac{\phi}{2}\Gamma_{\rho\phi}] \eta_0,
\end{equation}
where $\eta_0$ is a constant Majorana--Weyl spinor and $\Gamma_\rho,\,\Gamma_\phi$ are the flat gamma matrices in the plane generated by $\rho$ and $\phi$.
As pointed out above, since all the Killing spinors inevitably depend on $\phi$, acting with a TsT transformation breaks all supersymmetries.

\bigskip

In order to preserve supersymmetry under TsT transformations,  we must consider a more general situation involving more angular directions when performing the shift. We therefore consider a metric of the form
\begin{equation}
  \dd{ s^2} = -(\dd{ x^0})^2+ \sum_{i = 1}^{4} ( \dd{ \rho_{i}^{2}} + \rho_{i}^{2} \dd{\phi_{i}^{2}}) + \dd{ u^2}.
\end{equation}
In this frame, the Killing spinors are given by
\begin{equation}
  K = \prod_{i=1}^{4}\exp[\frac{\phi_{i}}{2}\Gamma_{\rho_{i}\phi_{i}}]\eta_{0}.
\end{equation}
We now consider a two-torus generated by \(u\) and a linear combination of the angles \(\phi_i\),
\begin{equation}
  \varphi = \sum_{i=1}^4 \frac{\varepsilon_i}{\norm{\varepsilon}} \phi_i,
\end{equation}
where the \(\varepsilon_i\) are real parameters and \(\norm{\varepsilon}^2 = \varepsilon_1^2 + \dots + \varepsilon_4^2\).
In the following we will identify the direction \(\varphi\) with the vector of parameters
\begin{equation}
  \varphi \mapsto (\varepsilon_1, \varepsilon_2, \varepsilon_3, \varepsilon_4) .
\end{equation}

The shift \( \varphi \to \varphi + \lambda \norm{\varepsilon} \tilde u\) is equivalent to the shifts \(\phi_i \to \phi_i + \lambda \varepsilon_i \tilde u\) that leave the three directions orthogonal to \(\varphi\) in the space generated by the \(\phi_i\) invariant.
To see this explicitly, let us introduce new angular variables \(\varphi_i\),
\begin{equation}
  \varphi_i = M_{ij} \phi_j,
\end{equation}
where \(M_{ij}\) is an orthogonal matrix (\(M^{-1} = M^t\)) and we fix
\begin{equation}
  M_{1j} = \frac{\varepsilon_j}{\norm{\varepsilon}}.
\end{equation}
Now consider the shifts \(\phi_j \to \phi_j + \lambda \varepsilon_j \tilde u\). The \(\varphi_i\) transform as
\begin{equation}
  \varphi_i = M_{ij} \phi_j \to M_{ij} \pqty{ \phi_j + \lambda \varepsilon_j \tilde u} = \varphi_i + \lambda M_{ij} \varepsilon_j \tilde u .
\end{equation}
But, \(\varepsilon_j = \norm{\varepsilon} M_{1j}\), so
\begin{equation}
  \varphi_i \to \varphi_i + \lambda \norm{\varepsilon} M_{ij} M_{1j} \tilde u = \varphi_i + \lambda \norm{\varepsilon} \delta_{1 i} \tilde u.
\end{equation}
Only the first variable \(\varphi_i = \varphi\) is shifted, while the others remain invariant.

Now let us see what happens to the Killing spinors.
First we rewrite \(K\) as a function of the \(\varphi_i\):
\begin{equation}
  K = \exp[ \sum_i \frac{\phi_i}{2} \Gamma_{\rho_i \phi_i}] \eta_0 = \exp[ \sum_{ij} \frac{1}{2} \varphi_j M_{ji} \Gamma_{\rho_i \phi_i}] \eta_0,
\end{equation}
where we have used the fact that the \(\Gamma_{\rho_i \phi_i}\) commute with each other and that
\begin{equation}
  \phi_i = (M^{-1})_{ij} \varphi_j = M_{ji} \varphi_j .
\end{equation}
What happens to \(K\) under the shift?
Since the shift acts on \(\varphi_1\) only and not on the other angles, we separate its contribution from the rest:
\begin{equation}
  \begin{aligned}
    K &= \exp[ \sum_{i} \frac{1}{2}M_{1i} \Gamma_{\rho_i \phi_i} + \sum_{\substack{i\\{j\neq 1}}} \frac{1}{2} \varphi_j M_{ji} \Gamma_{\rho_i \phi_i}] \eta_0 \\
    &\to \exp[ \sum_{i,j} \frac{1}{2} \varphi_j M_{ji} \Gamma_{\rho_i \phi_i} + \sum_{i} \frac{1}{2}\lambda \norm{\varepsilon} \tilde u M_{1i} \Gamma_{\rho_i \phi_i}  ] \eta_0 =  \exp[\frac{\lambda}{2} \tilde u \sum_i \varepsilon_i \Gamma_{\rho_i \phi_i}] K .
\end{aligned}
\end{equation}
As discussed in Section~\ref{sec:general}, only the Killing spinors that do \emph{not} depend on \(\tilde u\) survive the TsT transformation (or, rather, are compatible with the structure of the torus generated by \(u\) and \(\varphi\)).
This means that (part of the) supersymmetry can be preserved only if the operator \(\exp[\frac{\lambda}{2} \tilde u \sum_i \varepsilon_i \Gamma_{\rho_i \phi_i}] \) can be projected out.
Note that this is equivalent to requiring that the Killing spinor (before or after the shift) does not depend on \(\varphi\).

The analysis of the conditions under which supersymmetry is preserved was carried out in~\cite{arXiv:1204.4192,Lambert:2018cht}. The conditions on the $\varepsilon_{i}$ and the number of resulting supersymmetries are listed in Table~\ref{tab:SUSY}.
\begin{table}
  \centering
  \begin{tabular}{lc}
    \toprule
    conditions on $\varepsilon_{A}$ & unbroken SUSYs \\
    \midrule
    $\sum_{A=1}^{4}\varepsilon_{A} = 0$                                                & $4$        \\[5pt]
    $\sum_{A=1}^{3}\varepsilon_{A} = 0\,\&\, \varepsilon_{4}=0$                     & $8$              \\[5pt]
    $\varepsilon_{1}= \pm\varepsilon_{2}\, \&\, \varepsilon_{3} = \varepsilon_{4}=0$ & $16$                 \\[5pt]
    $\varepsilon_{1} = \pm\varepsilon_{2}\,\&\, \varepsilon_{3} = \pm\varepsilon_{4} $     & $8$               \\[5pt]
    $\varepsilon_{1} = \varepsilon_{2}=\varepsilon_{3}=\varepsilon_{4}$                & $12$                 \\[5pt]
    \bottomrule
  \end{tabular}
  \caption{Unbroken supersymmetries for different choices of the \(\varepsilon\) parameters.}
  \label{tab:SUSY}
\end{table}

\bigskip

The very same logic applies when we consider a torus generated by \(\varphi_1\) and \(\varphi_2\), which are both linear combinations of the angles \(\phi_i\).
The condition for preserving supersymmetry is that there are Killing spinors that do not depend on either \(\varphi_1\) or \(\varphi_2\).
This means that the coefficients in the two expansions \(\varphi_1 = \sum_i \varepsilon_i^{(1)} \phi_i\) and \(\varphi_2 = \sum_i \varepsilon_i^{(2)} \phi_i\) separately obey a condition in Table~\ref{tab:SUSY} \emph{and} that the corresponding projectors are not orthogonal.

\bigskip

We are now in the position of giving a general recipe for the Killing spinors preserved by TsT transformations in flat space.

\paragraph{The Recipe.}\

In order to perform a TsT transformation, we need two compact directions on which the group $SL(2,R)$ can act. Let us examine the nature of these periodic coordinates in which we perform the T--dualities.
If we act on a coordinate $u$ which corresponds to a flat direction compactified on an $S^1$, the natural $U(1)$ transformation which acts by shifting $u \to u + \alpha$ acts freely (\emph{i.e.} without fixed points) on the whole geometry.
If on the other hand the coordinate is an angular variable $\phi$ in a 2-plane parametrized by $\rho$ and $\phi$, the natural $U(1)$ transformation that shifts \(\phi \) has a fixed point at $\rho=0$.

We can therefore identify the following three possibilities depending on the type of the \(U(1)\)-action on the two directions of the torus:
\begin{enumerate}
\renewcommand{\labelenumi}{(\roman{enumi})}
\item both $U(1)$s act freely on the ten-dimensional background
\item only one $U(1)$ acts freely on the ten-dimensional background
\item the two $U(1)$s do not act freely on the ten-dimensional background.
\end{enumerate}
In all cases, we need to impose the condition that the Killing spinors do not depend on either of the generators of the torus.

Let us discuss these three possibilities in more detail:
\begin{enumerate}
\renewcommand{\labelenumi}{(\roman{enumi})}
\item $(u,v)_{\lambda}$\\
  The two $U(1)$ symmetries act freely.
  This means that the background metric contains the terms
  \begin{equation}
    \dd{u^{2}} + \dd{v^{2}}\,,
  \end{equation}
  and $\del_u$ and $\del_v$ are Killing vectors.
  Since the Killing spinor is completely independent of both coordinates throughout the whole TsT transformation $(u,v)_{\lambda}$, the supersymmetry is completely preserved.
The bi-vector that encodes the TsT transformation is given by
  \begin{equation}
    \Theta = \lambda \del_u \wedge \del_v = \Theta^{\mu\nu} \del_\mu \wedge \del_\nu,
  \end{equation}
  where
  \begin{equation}
    \Theta^{\mu\nu} =
    \begin{cases}
      \lambda & \text{if \(x^\mu = u\) and \(x^\nu = v\)} \\
      - \lambda & \text{if \(x^\mu = v\) and \(x^\nu = u\)} \\
      0 & \text{otherwise}.
    \end{cases}
  \end{equation}
  It is natural to associate this transformation to the canonical case of the \ac{nc} parameter \(\comm{\hat x^\mu}{\hat x^\nu} = i \Theta^{\mu \nu}\) of Section~\ref{sec:TsT-NC}.
\item $(u,\varphi)_{\lambda}$\\
  The next possibility is that only one of the \(U(1)\)s acts freely.
  The order of the variables is unimportant since, as we have already stressed, there is a geometric \(SL(2,\setR)\) action that exchanges them and commutes with the \(SL(2,\setR)\) that transforms \(\tau\).
  In general, the angle corresponding to the non-free action is a linear combination of the \(\phi_i\):
  \begin{equation}
    \varphi = \sum_i \frac{\varepsilon_i}{\norm{\varepsilon}} \phi_i.
  \end{equation}
Supersymmetry is thus preserved if the \(\varepsilon_i\) obey one of the conditions in Table~\ref{tab:SUSY}.

  We have already seen that the whole information about the TsT transformation is encoded in the bi-vector \(\Theta\) that in this case reads 
  \begin{equation}
  \Theta = \lambda \del_u \wedge \del_\varphi.
  \end{equation}
We can also extract the supersymmetry projector from it.
  If we rewrite the metric in flat coordinates we see that
  \begin{equation}
    \del_{\phi_i} = x^{2i -1 } \del_{x^{2i}} - x^{2i} \del_{x^{2i -1}}
  \end{equation}
  (from here on we will drop the \(x\) in the derivatives and write \(\del_{x^i} = \del_i\)).
  Then
  \begin{equation}\label{eq:thetauphi}
    \Theta = \lambda \del_u \wedge \del_\varphi = \lambda \del_u \wedge \sum_i \frac{\varepsilon_i}{\norm{\varepsilon}} \del_{\phi_i} = \lambda \del_u \wedge \sum_i \frac{\varepsilon_i}{\norm{\varepsilon}} \pqty{ x^{2i -1 } \del_{2i} - x^{2i} \del_{2i - 1}}.
  \end{equation}
 We want to construct an operator $P$ based on $\Theta$. For each term of the form $x^i\del_j-x^j\del_i$ in $\Theta$, we add a term $\Gamma_i\Gamma_j$ in this operator.
We thus associate to~\eqref{eq:thetauphi} the operator
  \begin{equation}
    P^\varphi = \sum_i \varepsilon_i \Gamma_{2i-1} \Gamma_{2i} .
  \end{equation}
  If the \(\epsilon_i\) obey a condition given in Table~\ref{tab:SUSY}, the operator \(P^\varphi\) is singular.
  \emph{We define the supersymmetry projector \(\proj{TsT}\) as the projector on the kernel of \(P^\varphi\)}.
  For example, it is easy to see that if the \(\varepsilon_i\) satisfy
  \begin{equation}
    \sum_{i=1}^N \varepsilon_i = 0 ,
  \end{equation}
  the corresponding projector takes the form
  \begin{equation}
    \proj{TsT} = \frac{1}{2^{N-1}}  \prod_{i=1}^{N-1} \pqty{ 1 - \Gamma_{2i-1} \Gamma_{2i} \Gamma_{2N-1} \Gamma_{2N}} .
  \end{equation}
  How do we relate the expression of \(\Theta \) to the non-commutativity description of Section~\ref{sec:TsT-NC}?
  If we rewrite the bi-vector as
  \begin{equation}
    \Theta = \lambda \del_u \wedge \del_\varphi = \Theta\indices{^\mu ^\nu _\rho} \, x^\rho \del_\mu \wedge \del_\nu,
  \end{equation}
  where
  \begin{equation}
    \Theta\indices{^\mu ^\nu _\rho} = \lambda \frac{ \varepsilon_i}{\norm{\varepsilon}} \times
    \begin{cases}
      1 & \text{if \(\mu = 2i -1\), \(\rho = 2i\), \(x^\nu = u\)}\\
      -1 & \text{if \(x^\mu = u\), \(\rho = 2i\), \(\nu = 2i - 1\)}\\
      -1 & \text{if \(\mu = 2i \), \(\rho = 2i - 1\), \(x^\nu = u\)}\\
      1 & \text{if \(x^\mu = u \), \(\rho = 2i - 1\), \(\nu = 2i\)}\\
      0 & \text{otherwise},
    \end{cases}
  \end{equation}
  it is natural to identify this transformation with the Lie algebra case \(\comm{\hat x^\mu}{\hat x^\nu} = i \Theta\indices{^\mu ^\nu_\rho} \hat x^\rho\).
\item $(\phi_{1}, \phi_{2})_{\lambda}$\\
  The last possibility is that both \(U(1)\)s have a fixed point.
  Then we can write them as
  \begin{align}
    \varphi_1 &= \sum_i \frac{\varepsilon_i^{(1)}}{\norm{\varepsilon^{(1)}}} \phi_i , & \varphi_2 &= \sum_i \frac{\varepsilon_i^{(2)}}{\norm{\varepsilon^{(2)}}} \phi_i  .
  \end{align}
  The bi-vector is now written in flat coordinates as
  \begin{equation}
    \Theta = \lambda \del_{\varphi_1} \wedge \del_{\varphi_2} = \lambda \sum_{ij} \frac{\varepsilon^{(1)}_i \varepsilon^{(2)}_j}{\norm{\varepsilon^{(1)}} \norm{\varepsilon^{(2)}}} \pqty{ x^{2i-1} \del_{2i} - x^{2i} \del_{2i-1}} \wedge \pqty{ x^{2j-1} \del_{2j} - x^{2j} \del_{2j-1}}  .
  \end{equation}
  In the same spirit as above, we can define two operators
  \begin{align}
    P^{\varphi_1} &= \sum_i \varepsilon^{(1)}_i \Gamma_{2i-1} \Gamma_{2i} , & P^{\varphi_2} &= \sum_i \varepsilon^{(2)}_i \Gamma_{2i-1} \Gamma_{2i} .
  \end{align}
  If both \(\varepsilon^{(1)}\) and \(\varepsilon^{(2)}\) obey a constraint given in Table~\ref{tab:SUSY}, we can define the supersymmetry projector \(\proj{TsT}\) as the projector on the intersection of the two kernels, \emph{i.e.} as the product of the two projectors on the two kernels.\\
  Once more we would like to associate the \(\Theta\) bi-vector to the \ac{nc} parameter.
  To do so, we can write
  \begin{equation}
      \Theta = \Theta\indices{^\mu ^\nu _\rho _\sigma} \, x^\rho x^\sigma \del_\mu \wedge \del_\nu,
    \end{equation}
  where
  \begin{equation}
    \Theta\indices{^\mu ^\nu _\rho _\sigma} = \lambda \frac{\varepsilon^{(1)}_i \varepsilon^{(2)}_j}{\norm{\varepsilon^{(1)}} \norm{\varepsilon^{(2)}}}  \times
    \begin{cases}
      1 & \text{if \(\mu = 2i\), \(\nu=2j\), \(\rho = 2i -1\), \(\sigma = 2j -1\)} \\
      -1 & \text{if \(\mu = 2j\), \(\nu=2i\), \(\rho = 2i -1\), \(\sigma = 2j -1\)} \\
      -1 & \text{if \(\mu = 2i\), \(\nu=2j-1\), \(\rho = 2i -1\), \(\sigma = 2j \)} \\
      1 & \text{if \(\mu = 2j-1\), \(\nu=2i\), \(\rho = 2i -1\), \(\sigma = 2j \)} \\
      -1  & \text{if \(\mu = 2i-1\), \(\nu=2j\), \(\rho = 2i\), \(\sigma = 2j -1\)} \\
      1  & \text{if \(\mu = 2j\), \(\nu=2i-1\), \(\rho = 2i\), \(\sigma = 2j -1\)} \\
      1  & \text{if \(\mu = 2i-1\), \(\nu=2j-1\), \(\rho = 2i\), \(\sigma = 2j\)} \\
      -1  & \text{if \(\mu = 2j-1\), \(\nu=2i-1\), \(\rho = 2i\), \(\sigma = 2j\)} \\
      0 & \text{otherwise}.
    \end{cases}
  \end{equation}
  It is natural to identify this case with the quantum space case of the \ac{nc} parameter in Section~\ref{sec:TsT-NC} \(\comm{\hat x^\mu}{\hat x^\nu} = i \Theta\indices{^\mu ^\nu _\rho _\sigma} x^\rho x^\sigma\).
\end{enumerate}

Once we know which Killing spinors survive the deformation, we can apply twice the formula in Eq.~\eqref{eq:Killing-after-TsT} and find an explicit expression for the Killing spinors in the deformed background that only depends on the bi-vector $\Theta$:
\begin{align}
  \epsilon_{L}^{\text{(fin)}} &= \proj{TsT} \epsilon_L^{\text{(in)}} , &
  \epsilon_{R}^{\text{(fin)}} &= \frac{1 + \frac{1}{2}\Theta^{\mu\nu}\hat{\Gamma}_{\mu}\hat{\Gamma}_{\nu}}{\sqrt{1+\frac{1}{2}\Theta^{\mu\nu}\Theta_{\mu\nu}}} \proj{TsT} \epsilon^{\text{(in)}}_{R}\,,
\end{align}
where the indices of the bi-vector are lowered with the initial undeformed metric.
Using the properties of the gamma matrices, this formula can be recast in exponential form (see also~\cite{arXiv:1204.4192}):
\begin{align}
  \epsilon_{L}^{\text{(fin)}} &= \proj{TsT} \epsilon_L^{\text{(in)}} , &
  \epsilon_{R}^{\text{(fin)}} &=  e^{\omega(\Theta) \, \Theta^{\mu\nu}\hat{\Gamma}_{\mu}\hat{\Gamma}_{\nu}}\proj{TsT} \epsilon^{\text{(in)}}_{R}\,,
\end{align}
where \(\omega(\Theta)\) is a normalization factor:
\begin{equation}
  \omega(\Theta) = \frac{\arctan(\sqrt{\frac{1}{2} \Theta^{\mu\nu} \Theta_{\mu\nu}})}{2 \sqrt{\frac{1}{2} \Theta^{\mu\nu} \Theta_{\mu\nu}}} .
\end{equation}

\subsection{Recipe for $\AdS{5} \times S^5$}
\label{sec:AdS5}

As in the previous section, we want to identify the preserved supersymmetries under TsT transformations on $\AdS{5} \times S^{5}$. 
We can use a similar recipe to the one for flat space by thinking of \(\AdS{5}\) (and \(S^5\)) as a surface in \(\setR^{2,4}\) (respectively \(\setR^6\)).

As we have stressed more than once, the choice of frame in which we write the Killing spinors is crucial.
We observe first of all that the Killing spinor equation for \(\AdS{5} \times S^5 \) in \tIIB string theory can be separated.
If we choose the representation of the ten-dimensional gamma matrices outlined in Appendix~\ref{app:Gamma}, a ten-dimensional spinor is written as
\begin{equation}
  \begin{aligned}
    \epsilon_{\text{10d}} &= 
    \begin{pmatrix}
      1\\0
    \end{pmatrix}
    \otimes \epsilon_{\AdS{5}} \otimes \epsilon_{S^{5}}\,,
  \end{aligned}
\end{equation}
where each five-dimensional spinor satisfies separately 
\begin{align}
  \label{eq:AdS5-gravitino}
  \AdS{5}:\ &D_{\mu}\epsilon_{\AdS{5}} = \frac{1}{2}\Gamma_{\mu}\epsilon_{\AdS{5}},\\
  \label{eq:S5-gravitino}
  S^{5}:\ &D_{j} \epsilon_{S^{5}} = \frac{i}{2}\Gamma_{j}\epsilon_{S^{5}}.
\end{align}
The embedding of the five-dimensional Gamma-matrices in ten dimensions is obtained by multiplying each Gamma matrix by $i\hat{\gamma}= -\sigma_{2} \otimes \mathbf{1}_{4\times4} \otimes \mathbf{1}_{4\times4}$.

There are various ways of parametrizing $\AdS{5}\times S^{5}$, which each have their use depending on the nature of the TsT transformation. We give a list of Killing spinors in different coordinate systems for $\AdS{5}$ and $S^{5}$, respectively, and then propose a recipe for identifying the preserved supersymmetries on deformed $\AdS{5}\times S^{5}$ backgrounds, relating them to the bi-vector defined in Section~\ref{sec:R-matrix}. 
\paragraph{Killing spinors in $\AdS{5}$.}

Let us start with the $\AdS{5}$ side.
$\AdS{5}$ space is defined as the locus in \(\setR^{2,4}\) that obeys the equation
\begin{equation}\label{eq:horos-coord}
  - (T^1)^2 - (T^2)^2 + (X^1)^2 + (X^2)^2 + (X^3)^2 + (X^4)^2 = -1 ,
\end{equation}
where the horospherical coordinate system is defined as
\begin{equation}
  \begin{cases}
    T^1 = \frac{e^{-r}}{2} \pqty{ 1 + e^{2r} \pqty{ 1 - (x^0)^2 + (x^1)^2 + (x^2)^2 + (x^3)^2}} ,\\
    T^2 = e^r x^0,\\
    x^1 = e^r x^1, \\
    x^2 = e^r x^2, \\
    x^3 = e^r x^3 ,\\
    x^4 = \frac{e^{-r}}{2} \pqty{ 1 - e^{2r} \pqty{ 1 + (x^0)^2 - (x^1)^2 - (x^2)^2 - (x^3)^2}}  .
  \end{cases}
\end{equation}
The corresponding metric is given by
\begin{equation}
  \dd{ s^{2}}_{\AdS{5}} = e^{2r}\eta_{\alpha\beta} \dd{ x^{\alpha}} \dd{ x^{\beta}} + \dd{ r^{2}}\,,
\end{equation}
where $\eta_{\alpha\beta} = {\rm diag}(-,+,+,+)$ for $\alpha,\beta = 0,1,2,3$.
In this frame, the vanishing of the gravitino variation can be separated into an equation for $x^{\alpha}$ and an equation for $r$:
\begin{equation}
\begin{aligned}
\partial_{\alpha}\epsilon &= \frac{1}{2}\Gamma_{\alpha}\left(1 - \Gamma_{r}\right)\epsilon\,,\\
\partial_{r}\epsilon &=\frac{1}{2}\Gamma_{r}\epsilon\,.
\end{aligned}
\end{equation}
The result is that the Killing spinors have the form~\cite{Lu:1996rhb}
\begin{equation}
  \label{eq:Killing-spinors-horospherical-AdS}
  \boxed{\epsilon_{\AdS{5}} = e^{\frac{1}{2}r\Gamma_{r}}\left(1+\frac{1}{2}x^{\alpha}\Gamma_{\alpha}(1-\Gamma_{r})\right)\epsilon_{0}\,,}
\end{equation}
where $\epsilon_{0}$ is a constant spinor in five dimensions.
The important observation for our purposes is that only half of the spinors depend on the coordinates \(x^\alpha\) and they can be projected out with the projector
\begin{equation}
  \proj{r} = \frac{1}{2} \pqty{ 1 + \Gamma_r}.
\end{equation}

\medskip

Another useful parametrization in which the \(O(2)^3\) maximal torus of the \(SO(2,4)\) symmetry group is manifest is obtained if, instead of light-cone coordinates, we introduce three pairs of complex coordinates $\{\tau_{i}\}_{i=1,2,3}$.
Then the action of the maximal torus is diagonalized in terms of the three phases \(\varphi_i\):
\begin{equation}
\begin{cases}
\tau_{1} = t_{1} + i t_{2} =  \sinh(\theta_{1}) \cos(\theta_{2}) e^{i\phi_{1}}\,,\\
\tau_{2} = x_{1} + i x_{2} = \sinh(\theta_{1}) \sin( \theta_{2}) e^{i\phi_{2}}\,,\\
\tau_{3} = x_{3} + i x_{4}  =  \cosh(\theta_{1}) e^{i\phi_{3}}\,.
\end{cases}
\end{equation}
The corresponding metric is given by
\begin{multline}
  \dd{s_{\AdS{5}}}^{2} = \dd{\theta_{1}}^{2} + \sinh[2](\theta_{1}) \dd{\theta_{2}}^{2}   + \sinh^{2} (\theta_{1}) \cos[2](\theta_{2}) \dd{\phi_1}^{2} \\+ \sinh[2](\theta_{1}) \sin[2](\theta_{2}) \dd{\phi_2}^{2} - \cosh[2](\theta_{1}) \dd{\phi_3}^{2} \,,
\end{multline}
and the corresponding Killing spinor is given by
\begin{equation}
  \boxed{\epsilon = e^{\frac{1}{2}\theta_{1}\Gamma_{\theta_{1}}}e^{\frac{1}{2}\theta_{2}\Gamma_{\theta_{1}\theta_{2}}} e^{\frac{i}{2}\phi_3\Gamma_{\phi_3}} e^{\frac{i}{2}\phi_1 \Gamma_{\theta_{1} \phi_1 }} e^{\frac{1}{2}\phi_2\Gamma_{\theta_{2}\phi_2}}\epsilon_{0}}\,.
\end{equation}
Unsurprisingly, the form of the Killing spinor is reminiscent of the flat space expression in polar coordinates in Eq.\eqref{eq:flat-Killing-spinor1}.
This will allow us to introduce a similar recipe for the preserved Killing spinors in terms of projectors.

\paragraph{Killing spinors in $S_{5}$.}

Let us now consider the $S^{5}$ part.
The most useful parametrization is similar to the last one that we have seen for \(\AdS{5}\) and makes the \(O(2)^3\) maximal torus of \(SO(6)\) manifest.
We write the five-sphere as a surface in \(\setR^6\):
\begin{equation}
  (X^1)^2 + (X^2)^2 + (X^3)^2 + (X^4)^2 + (X^5)^2 + (X^6)^2 = 1,
\end{equation}
which we parametrize explicitly as
\begin{equation}
  \begin{cases}
    X^1 + i X^2  = \rho_1 e^{i \phi_4} = \sin( \theta_4) \cos(\theta_5) e^{i \phi_4}, \\
    X^3 + i X^4  = \rho_2 e^{i \phi_5} = \sin( \theta_4) \sin(\theta_5)  e^{i \phi_5}, \\
    X^5 + i X^6  = \rho_3 e^{i \phi_6} = \cos( \theta_4) e^{i \phi_6}.     
  \end{cases}
\end{equation}
The metric is given by
\begin{multline}
\label{eq:metric-S5}
  \dd s_{S^{5}}^{2} =%
   \dd{\theta_4}^{2} + \sin[2](\theta_4) \dd{\theta_5}^{2}  + \sin[2](\theta_4)  \cos[2](\theta_5) \dd{\phi_4}^{2} \\
  + \sin[2](\theta_4) \sin[2](\theta_5) \dd{\phi_5}^{2} + \cos[2]( \theta_4) \dd{\phi_6}^{2} \,.
\end{multline}
It is straightforward to compute the solution of the gravitino equation~(\ref{eq:S5-gravitino}), which is given by\footnote{See Appendix~\ref{sec:spinor-detail} for details.}
\begin{equation}
  \boxed{\epsilon_{S^{5}} = e^{\frac{i}{2}\theta_4 \Gamma_{\theta_4}} e^{\frac{\theta_5}{2}\Gamma_{\theta_4\theta_5}} e^{\frac{i}{2}\phi_6\Gamma_{\phi_6}} e^{\frac{\phi_4}{2}\Gamma_{\theta_4\phi_4}} e^{\frac{\phi_5}{2}\Gamma_{\theta_5\phi_5}}\tilde{\epsilon}_{0}}\,.
\end{equation}
Once more we find a form that is close to the flat space Killing spinors of Eq.~(\ref{eq:flat-Killing-spinor1}).
We are now ready to give the recipe.

\paragraph{Recipe for $\AdS{5} \times S^{5}$. }

In the case of a TsT transformation of flat space we have distinguished between freely-acting and non-freely-acting \(U(1)\) symmetries.
In the \(\AdS{5} \times S^5\) case the distinction is more delicate because even the analogue of the freely-acting \(U(1)\) has a fixed point which, as we will shortly see, means that supersymmetry is always partially broken.
For our purposes it is however still convenient to introduce a distinction, this time based on the action of the \(U(1)\)s on the embedding flat space \(\setR^{2,4} \times \setR^6\).

We have again three possibilities:
\begin{enumerate}
\renewcommand{\labelenumi}{(\roman{enumi})}
\item $(u,v)_{\lambda}$\\
  If both \(U(1)\)s act freely on \(\setR^{2,4}\), the most convenient frame is given by horospherical coordinates on \(\AdS{5}\). The explicit dependence on \(u\) and \(v\) (which are identified with two \(x^\alpha\) directions) is eliminated by the projector
  \begin{equation}
    \proj{r} = \frac{1}{2 } \pqty{ 1 + i\hat{\gamma}\hat{\Gamma}_r}
  \end{equation}
  which projects out half of the Killing spinors.%
\item $(u, \varphi)_\lambda$\\
  If one of the \(U(1)\)s acts freely on \(\setR^{2,4}\) and the other acts on \(\setR^6\) with a fixed point, the \(u\)-dependence is eliminated by the projector \(\proj{r}\), while the \(\varphi\)-dependence can be projected out only if \(\varphi\) is a linear combination of the three angles \(\phi_4, \phi_5, \phi_6\),
  \begin{equation}
    \varphi = \sum_{i = 4}^6 \frac{\varepsilon_i}{\norm{\varepsilon}} \phi_i,
  \end{equation}
  and the \(\varepsilon_i\) satisfy one of the conditions in Table~\ref{tab:SUSY}.
  In this case, the construction of the corresponding projector is similar in spirit to the one in flat space.
  We start from the bi-vector
  \begin{equation}
    \Theta = \lambda \del_u \wedge \del_\varphi = \frac{\lambda}{\norm{\varepsilon}} \del_u \wedge \sum_{i=4}^6 \varepsilon_i \del_{\phi_i}
  \end{equation}
  and we introduce the operator \(P^\varphi\) which contains a term \(\Gamma_{\theta_4 \phi_4}\), \(\Gamma_{\theta_5 \phi_5}\) or \(i \Gamma_{\phi_6}\) if respectively \(\del_{\phi_4}\), \(\del_{\phi_5}\) or \(\del_{\phi_6}\) appear in \(\Theta\).
  The projector \(\proj{TsT}\) is then given by
  \begin{equation}
    \proj{TsT} = \proj{r} \proj{$\varphi$},
  \end{equation}
  where \(\proj{$\varphi$}\) projects on the kernel of \(P^\varphi\).
\item $(\varphi_1, \varphi_2)_\lambda$\\
  In the last case both \(U(1)\)s act with a fixed point on \(\setR^{2,4}\) and/or \(\setR^6\).
  This case is directly analogous to the one discussed for flat space.
  Supersymmetry is preserved if both angles are linear combinations of the \(\phi_i\) and obey a constraint in Table~\ref{tab:SUSY}.
  The projector is again obtained by starting from the operators \(P^{\varphi_1}\) and \(P^{\varphi_2}\) which contain a term \(\Gamma_{\theta_1 \phi_1}\), \(\Gamma_{\theta_2 \phi_2}\) or \(i \Gamma_{\phi_3}\) if respectively \(\del_{\phi_1}\), \(\del_{\phi_2}\) or \(\del_{\phi_3}\) appear in \(\Theta\).
  The final projector is the product of the two projectors on the kernels of the operators \(P^{\varphi_1}\) and \(P^{\varphi_2}\),
  \begin{equation}
    \proj{TsT} = \proj{$\varphi_1$} \proj{$\varphi_2$}.
  \end{equation}
\end{enumerate}

\section{Examples}
\label{sec:examples}

In this section, we give explicit expressions for the Killing spinors of a variety of deformations both of flat space and AdS-space which have appeared in the literature in the contexts of integrable \acp{ybd} and deformed supersymmetric gauge theories.
For each deformation we indicate the corresponding $r$-matrix, expressed in terms of the generators of the relevant symmetry algebra, where \(p_i\) is the momentum generator and \(n_{ij}\) is a rotation in the plane generated by \(x_i\) and \(x_j\).

\subsection{Killing spinors for deformations of 10d flat space}
\label{sec:example-flat-space}
\subsubsection{Seiberg--Witten: $r = p_1 \wedge p_2$} \
If we start from flat spacetime in rectangular coordinates $\{x^{\mu}\}_{\mu=0,1,...,9}$, the Killing spinor remains constant under a TsT transformation parametrized by $(x^{1},x^{2})_{\lambda}$. Therefore, all supersymmetries are preserved. After applying $(x^{1},x^{2})_{\lambda}$, the background is given by
\begin{equation}
  \begin{aligned}
    \dd{s}^{2} &= -(\dd{x^0})^{2} + \frac{(\dd{x^1})^{2} + (\dd{x^2})^{2}}{1+\lambda^{2}} + \sum_{m=3}^9 (\dd{x^m})^{2} ,\\
    e^{-2\Phi} &= 1+\lambda^{2},\\
    B_{2} &= \frac{\lambda}{1+\lambda^{2}} \dd {x^1} \wedge \dd{x^{2}}\,.
  \end{aligned}
\end{equation}
The Seiberg--Witten map tells us that this translates to a non-commutative gauge theory on flat space with \ac{nc} parameter 
\begin{equation}
\Theta = -\lambda \partial_{1} \wedge \partial_{2}\,.
\end{equation}
This deformation corresponds exactly to the one described first in~\cite{hep-th/9908142} and the \ac{ybd} 
with classical $r$-matrix~\cite{Matsumoto:2015ypa}
\begin{equation}
r = p_1 \wedge p_2.
\end{equation}

\subsubsection{$\Omega$--deformation with $\varepsilon_1 = \varepsilon_2$: $r = p_1 \wedge (n_{23} + n_{45})$} \
Let us parametrize two 2-planes in flat space via the polar coordinates $(\rho_{1},\phi_{1})$ and $(\rho_{2},\phi_{2})$:
\begin{equation}
\dd{s}^{2} = -(\dd{x^{0}})^{2} + (\dd{x^1})^{2}+ \dd{\rho_{1}}^{2} + \rho_{1}^{2} \dd{\phi_{1}}^{2} + \dd{\rho_{2}}^{2} + \rho_{2}^{2} \dd{\phi_{2}}^{2} + \sum_{m=6}^{9}(\dd{x^{m}})^{2}\,,
\end{equation}
which implies that the Killing spinor is the same as the one given in Eq.~\eqref{eq:flat-Killing-spinor1}.
We choose $x^{1}$ to be a circle fiber direction. We rewrite the two angles $\phi_{1}$ and $\phi_{2}$ as 
\begin{equation}
\phi_{+} = \frac{\phi_{1}+\phi_{2}}{2}\,,\qquad \phi_{-} = \frac{\phi_{1}-\phi_{2}}{2}\,.
\end{equation}
The $\Omega$ deformation~\cite{Nekrasov:2002qd} with deformation parameters $\varepsilon_1 = \varepsilon_2$ corresponds to the TsT transformation $(x^{1}, \phi_{+})_{\lambda}$. 
Both angles $(\phi_{1}, \phi_{2})$ receive a monodromy with respect to $x^{1}$, which is nothing but the shift transformation by $\frac{\lambda}{2} x^{1}$. Lastly, we perform a T-duality in $x^{1}$. Note that $\phi_{-}$ remains unchanged under this TsT transformation as discussed in Section~\ref{sec:flat}. 
We denote the directions $\phi_{+}$ and $\phi_{-}$  by
\begin{align}
  \phi_{+} &= (1,1,0)\,, & \phi_{-} &= (1,-1,0)\,.
\end{align}
The above expression shows that both $(x^{1}, \phi_{\pm})_{\lambda}$ are supersymmetric choices. $(x^{1}, \phi_{+})_{\lambda}$ preserves half of the supersymmetry via the projector associated to $\phi_{+}$,
\begin{equation}
\proj{$\phi_{+}$}  = \frac{1}{2}(1 + \hat{\Gamma}_{\rho_{1}\phi_{1} \rho_{2}\phi_{2}})\,,
\end{equation}
which is encoded in the Killing spinor as follows:
\begin{equation}
\tilde{\epsilon} = e^{\phi_{-}\hat{\Gamma}_{\rho_{1}\phi_{1}}}\proj{$\phi_{+}$}\epsilon_{0}.
\end{equation}
The resulting background after $(x^{1},\phi_{+})_{\lambda}$ can be rewritten in terms of the original angles $(\phi_{1},\phi_{2})$ as
\begin{equation}
  \begin{aligned}
    \dd{s}^{2} ={}& -(\dd{x^{0}})^{2} + \frac{(\dd{x^{1}})^{2}}{1+ \lambda^{2}(\rho_{1}^{2} + \rho_{2}^{2})} + \dd{\rho_{1}}^{2} + \frac{1+ \lambda^{2}\rho_{2}^{2}}{1+ \lambda^{2}(\rho_{1}^{2} + \rho_{2}^{2})}\rho_{1}^{2} \dd{\phi_{1}}^{2}\\
    & + \dd{\rho_{2}}^{2} + \frac{1+ \lambda^{2}\rho_{1}^{2}}{1+ \lambda^{2}(\rho_{1}^{2} + \rho_{2}^{2})}\rho_{2}^{2} \dd{\phi_{2}}^{2} - \frac{2 \lambda^{2}\rho_{1}^{2}\rho_{2}^{2}}{1+ \lambda^{2}(\rho_{1}^{2} + \rho_{2}^{2})} \dd{\phi_{1}} \dd{\phi_{2}}\\
    & + \sum_{m=6}^{9}(\dd{x_{m}})^{2},\\
    e^{-2\Phi} ={}& 1+ \lambda^{2}(\rho_{1}^{2}+\rho_{2}^{2}),\\
    B_{2} ={}& \frac{\lambda}{1+\lambda^{2}(\rho_{1}^{2}+\rho_{2}^{2})} \dd{x_{1}} \wedge \pqty{\rho_{1}^{2} \dd{\phi_{1}} + \rho_{2}^{2} \dd{\phi_{2}}}\,.
  \end{aligned}
\end{equation}
The above background corresponds to the fluxtrap background introduced in~\cite{arXiv:1106.0279, arXiv:1309.7350} which realizes the $\Omega$-deformation via a deformed string theory background.
Generalizing the Seiberg--Witten map of the previous example, the background is flat with a bi-vector given by
\begin{equation}
\Theta = -\lambda \partial_{x^{1}} \wedge ( \partial_{\phi_{1}} + \partial_{\phi_{2}} ) =  - \lambda \partial_{x^{1}} \wedge \partial_{\phi_{+}}\,. 
\end{equation} 
Rewriting the polar coordinates $(\rho_{1},\phi_{1})$ and $(\rho_{2},\phi_{2})$ in terms of $(x^{2},x^{3})$ and $(x^{4},x^{5})$, we note that the bi-vector is of the Lie-algebra type:
\begin{equation}
\begin{aligned}
\Theta &= -\lambda  \partial_{x^{1}} \wedge \biggl[(x^{2}\partial_{x^{3}} - x^{3} \partial_{x^{2}}) + (x^{4}\partial_{x^{5}} - x^{5}\partial_{x^{4}})\biggr]\\
&=-\lambda \epsilon_{1jk}x^{j}\partial_{x^{1}} \wedge \partial_{x^{k}}\,,
\end{aligned}
\end{equation}
where $\epsilon_{123} = \epsilon_{145} = 1$.
As far as we know, this background has not been discussed as a \ac{ybd} so far.
In principle, by performing a \ac{ybd} with the classical $r$-matrix
\begin{eqnarray}
r = p_1 \wedge (n_{23} + n_{45}), 
\end{eqnarray}
the deformed background should be reproduced.
The typical interpretation in terms of a \ac{ybd} works when the $r$-matrix lives completely in four spacetime dimensions~\cite{Matsumoto:2015ypa, arXiv:1510.03083}, while the configuration derived here is more general, since it involves at least one extra direction.
However, our result is reproduced if the \ac{ybd} is understood as a $\beta$-transformation~\cite{arXiv:1705.07116, arXiv:1708.03163, 
arXiv:1803.05903}.

The relationship between the $\Omega$-deformation and non-commutativity has already been observed in the past~\cite{Aganagic:2003qj, Dijkgraaf:2008ua, Mironov:2009dv, arXiv:1204.4192}. Looking at the $\Omega$-deformation as a TsT transformation gives rise to a natural interpretation in terms of integrable deformations and their related non-commutativity parameter.
It would be interesting to explore this observation in more detail. 
\begin{equation}
  \label{eq:Killing-after-TsTflat2}
  \begin{cases}
    {\epsilon^{\text{(fin)}}}_L = \epsilon^{\text{(in)}}_L , \\
    {\epsilon^{\text{(fin)}}}_R = \frac{1}{\Delta}( 1- \lambda (\rho_{1}\hat{\Gamma}_{x^{1}\phi_{1}}+\rho_{2}\hat{\Gamma}_{x^{1}\phi_{2}})) {\epsilon}^{\text{(in)}}_R\,,
  \end{cases}
\end{equation}
where $\Delta = \sqrt{1+ \lambda^{2}(\rho_{1}^{2}+\rho_{2}^{2})}$ and 
\begin{equation}
\epsilon^{\text{(in)}} = \epsilon^{\text{(in)}}_{L}+\epsilon^{\text{(in)}}_{R} =  e^{\frac{\phi_{1}-\phi_{2}}{2}\hat{\Gamma}_{\rho_{1}\phi_{1}}}\proj{$\phi_{+}$}\epsilon_{0}\,.
\end{equation}
\subsubsection{Lunin--Maldacena-like deformation of flat space: $r = (n_{12} + n_{34}) \wedge (n_{34} + n_{56})$}
\label{sec:two-U1-fixed-points}

The situation becomes more complicated if we use three sets of polar coordinates,
\begin{equation}
  \dd{s}^{2} = -(\dd{x^{0}})^{2} + (\dd{x^{1}})^{2} + \sum_{i=1}^{3}( \dd{\rho_{i}}^{2} + \rho_{i}^{2} \dd{\phi_{i}}^{2}) + (\dd{x^{8}})^{2} + (\dd{x^{9}})^{2}, 
\end{equation}
and use the following angles:
\begin{equation}
\phi_{1} = \psi - \varphi_{1}\,,\qquad \phi_{2} = \psi + \varphi_{1} + \varphi_{2}\,,\qquad \phi_{3} = \psi - \varphi_{2}\,.
\end{equation}
Using the recipe, one can read off
\begin{equation}
\varphi_{1} = (-1,1,0)\,,\qquad \varphi_{2} = (0,1,-1)\,,\qquad \psi = (1,1,1)\,,
\end{equation}
which implies that a TsT transformation involving $\psi$ does not preserve any supersymmetry whereas $(\varphi_{1}, \varphi_{2})_{\lambda}$ results in a supersymmetric configuration.
The corresponding  Killing spinor contains the following projectors: 
\begin{equation}
\begin{aligned}
  \proj{$\varphi_{1}$} &= \frac{1}{2}(1-\hat{\Gamma}_{\rho_{1}\phi_{1}\rho_{2}\phi_{2}}) \,,\\
  \proj{$\varphi_{2}$} &= \frac{1}{2}(1- \hat{\Gamma}_{\rho_{2}\phi_{2}\rho_{3}\phi_{3}})\,.
\end{aligned}
\end{equation}
Therefore, only  one quarter of the supersymmetry is preserved.  The deformed background has the form
\begin{equation}
\begin{aligned}
\dd{s}^{2} ={}& -(\dd{x^{0}})^{2} + (\dd{x^{1}})^{2} + \dd{\rho_{1}}^{2} + \dd{\rho_{2}}^{2} + \dd{\rho_{3}}^{2} + (\dd{x^{8}})^{2} + (\dd{x^{9}})^{2}\\ 
&+ \frac{1}{\Delta^{2}}(\rho_{1}^{2}( \dd{\psi} - \dd{\varphi_{1}})^{2} + \rho_{2}^{2} ( \dd{\psi} + d{\varphi_{1}} + \dd{\varphi_{2}})^{2} + \rho_{3}^{2}( \dd{\psi} - \dd{\varphi_{2}})^{2} + 9\lambda^{2}\rho_{1}^{2}\rho_{2}^{2}\rho_{3}^{2} \dd{\psi}^{2})\,, \\
e^{-2\Phi} ={}&  \Delta^{2}\,,\\
B_{2} ={}& \frac{\Delta^{2}-1}{\lambda \Delta^{2}} \dd{\varphi_{1}} \wedge \dd{\varphi_{2}} - \frac{\Delta^{2} - 1}{\lambda\Delta^{2}}( \dd{\varphi_{1}} - \dd{\varphi_{2}}) \wedge \dd{\psi} + \frac{3 \lambda }{\Delta^{2}}(\rho_{1}^{2}\rho_{2}^{2} \dd{\varphi_{1}} - \rho_{2}^{2}\rho_{3}^{2} \dd{\varphi_{2}}) \wedge \dd{\psi} \,,
\end{aligned}
\end{equation}
where $\Delta^{2} = 1+ \lambda^{2} (\rho_{1}^{2}\rho_{2}^{2} + \rho_{2}^{2}\rho_{3}^{2} + \rho_{3}^{2}\rho_{1}^{2})$\,. 
However, the bi-vector $\Theta$ is rather simple, and can be expressed in different ways:
\begin{equation}
\begin{aligned}
\Theta &= - \lambda \partial_{\varphi_{1}} \wedge \partial_{\varphi_{2}}\\ 
&= \lambda (\partial_{\phi_{1}} \wedge \partial_{\phi_{2}} + \partial_{\phi_{2}} \wedge \partial_{\phi_{3}} + \partial_{\phi_{3}} \wedge \partial_{\phi_{1}})\\
&=\lambda (\epsilon_{1ij}\epsilon_{3mn} x^{i}x^{n}\partial_{x^{j}}\wedge \partial_{x^{n}} + \epsilon_{3mn}\epsilon_{5pq}x^{m}x^{p} \partial_{x^{n}} \wedge \partial_{x^{q}} + \epsilon_{5pq}\epsilon_{1ij}x^{p}x^{i}\partial_{x^{p}}\wedge \partial_{x^{j}})\,,
\end{aligned}
\end{equation}
where the anti-symmetric tensor $\epsilon_{123} = \epsilon_{345} = \epsilon_{567} = 1$. 

Thus the associated classical $r$-matrix is given by 
\begin{eqnarray}
r = (n_{12} + n_{34}) \wedge (n_{34} + n_{56}). 
\end{eqnarray} 
This is intrinsic to six dimensions, and hence due to the same reasoning as in the previous example, 
this background has not been considered 
as a \ac{ybd}\footnote{The four-dimensional analogue was discussed in~\cite{hep-th/0509036}. It can be described as a \ac{ybd}~\cite{Matsumoto:2015ypa}.}.  It can be derived by employing a $\beta$-transformation 
procedure~\cite{arXiv:1705.07116, arXiv:1708.03163, arXiv:1803.05903}.

This example is the analogue of the Lunin--Maldacena deformation (discussed in Sec.~\ref{sec:ML}) in flat space.
The surviving Killing spinors after the TsT transformation are given by
\begin{equation}
  \label{eq:Killing-after-TsTflat3}
  \begin{cases}
    {\epsilon^{\text{(fin)}}}_L = \epsilon^{\text{(in)}}_L , \\
    {\epsilon^{\text{(fin)}}}_R = \frac{1}{\Delta}\biggl( 1+ \lambda (\rho_{1}\rho_{2}\hat{\Gamma}_{\phi_{1}\phi_{2}}+\rho_{2}\rho_{3}\hat{\Gamma}_{\phi_{2}\phi_{3}}+\rho_{3}\rho_{1}\hat{\Gamma}_{\phi_{3}\phi_{1}})\biggr){\epsilon}^{\text{(in)}}_R\,,
  \end{cases}
\end{equation}
where
\begin{equation}
\epsilon^{\text{(in)}} = \epsilon^{\text{(in)}}_{L}+\epsilon^{\text{(in)}}_{R} =  e^{\frac{\psi}{2}(\hat{\Gamma}_{\rho_{1}\phi_{1}}+\hat{\Gamma}_{\rho_{2}\phi_{2}}+\hat{\Gamma}_{\rho_{3}\phi_{3}})}\proj{$\varphi_{1}$}\proj{$\varphi_{2}$}\epsilon_{0}.
\end{equation}
\subsection{Killing spinors for deformations of $\AdS{5} \times S^5$}
\label{sec:example-AdS5-S5}

We now move on to deformations of $\AdS{5} \times S^5$ which are related to the holographic descriptions of deformations of four-dimensional gauge theories.

\subsubsection{Hashimoto--Izhaki--Maldacena--Russo: $r = p_1 \wedge p_2$} 

As a first example, we consider a TsT transformation with both $U(1)$s acting freely on the embedding space $\setR^{2,4}$. This background was first discussed in~\cite{hep-th/9907166,hep-th/9908134}. This can be seen as a \ac{ybd} with classical $r$-matrix~\cite{arXiv:1404.3657} 
\begin{eqnarray}
r = p_1 \wedge p_2. 
\end{eqnarray}

Starting from $\AdS{5}$ parametrized in horospherical coordinates $(x^{\alpha}, r)$, we can perform any TsT-transformation with $x^{\alpha}$, say, $(x^{1},x^{2})_{\lambda}$, preserving half of the supersymmetry in the end. The bulk fields transform using $\Delta^{2} = 1+\lambda^{2}e^{4r}$ into
\begin{equation}
\begin{aligned}
\dd{s}^{2} &= \dd{r}^{2} + e^{2r} \left(-(\dd{x^{0}})^{2} + \frac{(\dd{x^{1}})^{2} + (\dd{x^{2}})^{2}}{\Delta^{2}} + (\dd{x^{3}})^{2}\right) + \dd{s}_{S^{5}}^{2}\,,\\
B_{2} &= \frac{\lambda e^{4r}}{\Delta^{2}} \dd{x^{1}}\wedge \dd{x^{2}}\,,\\
e^{-2\Phi} &= \Delta^{2}\,,\\
C_{2} &=-\lambda e^{4r} \dd{x^{0}} \wedge \dd{x^{3}}\,,\\
C_{4} &= \frac{e^{4r}}{\Delta^{2}} \dd{x^{0}} \wedge \dd{x^{1}} \wedge \dd{x^{2}} \wedge \dd{x^{3}} + \cos(\theta_{1})\sin(\theta_{1})\sin[4](\theta_{2}) \dd{\theta_{1}} \wedge \dd{\phi_{1}} \wedge \dd{\phi_{2}} \wedge \dd{\phi_{3}}\,.
\end{aligned}
\end{equation}
The corresponding bi-vector is simply given by
\begin{equation}
\Theta = -\lambda \partial_{x^{1}} \wedge \partial_{x^{2}}\,.
\end{equation}
Recall that in this frame, the Killing spinor on $\AdS{5}$ is given by
\begin{equation}
\epsilon_{\mathrm{AdS}_{5}} = e^{\frac{1}{2}r\Gamma_{r}}\left(1 + \frac{1}{2}x^{\alpha}\Gamma_{\alpha}(1-\Gamma_{r})\right)\epsilon_{0}\,.
\end{equation}
The $x^{\alpha}$-dependence can be removed from the spinor via the projector
\begin{equation}
\proj{r} = \frac{1}{2}(1 + \Gamma_{r})\,.
\end{equation}
Hence, starting from 
\begin{equation}
\begin{aligned}
\epsilon_{{\rm \AdS{5}}} &= e^{\frac{1}{2}r\Gamma_{r}}\left(1 + \frac{1}{2}x^{\alpha}\Gamma_{\alpha}(1-\Gamma_{r})\right) \proj{r} \epsilon_{0}\\
&=e^{\frac{1}{2}r\Gamma_{r}}\proj{r}\epsilon_{0}\,,
\end{aligned}
\end{equation}
the final result for the Killing spinors is
\begin{equation}
  \label{eq:Killing-after-MR}
  \begin{cases}
    {\epsilon^{\text{(fin)}}}_L = \epsilon^{\text{(in)}}_L \\
    {\epsilon^{\text{(fin)}}}_R = \frac{1}{\Delta}(1-\lambda e^{2r}\hat{\Gamma}_{x^{1}x^{2}}){\epsilon^{\text{(in)}}}_R ,
  \end{cases}
\end{equation}
with
\begin{equation}
\epsilon^{\text{(in)}} = \epsilon^{\text{(in)}}_{L}+\epsilon^{\text{(in)}}_{R}=e^{\frac{i}{2}\hat{\gamma}\hat{\Gamma}_{r}}e^{-\frac{1}{2}\theta_{2}\hat{\gamma}\hat{\Gamma}_{\theta_{2}}}e^{-\frac{1}{2}\phi_{3}\hat{\gamma}\hat{\Gamma}_{\phi_{3}}}e^{\frac{\theta_{1}}{2}\hat{\Gamma}_{\theta_{2}\theta_{1}}}e^{\frac{\phi_{1}}{2}\hat{\Gamma}_{\theta_{2}\phi_{1}}}e^{\frac{\phi_{2}}{2}\hat{\Gamma}_{\theta_{1}\phi_{2}}}\, \proj{$r$}\epsilon_{0}\,,
\end{equation}
where the projector $\proj{r}$ is uplifted to the ten-dimensional expression.

\subsubsection{Lunin--Maldacena: $r = (n_{12} + n_{34}) \wedge (n_{34} + n_{56})$}
\label{sec:ML}

The next case we consider is a TsT transformation which uses two angular variables in $S^5$, which means that neither of the $U(1)$s acts freely in the embedding space.
This background appeared first in~\cite{hep-th/0502086}. This case can be understood as a \ac{ybd} with classical $r$-matrix~\cite{arXiv:1404.1838} 
\begin{eqnarray}
r = (n_{12} + n_{34}) \wedge (n_{34} + n_{56}). 
\end{eqnarray}
We start from the metric given by
\begin{equation}
\begin{aligned}\label{eq:LM S5} 
 \dd{s}^{2}={}& \dd{s}^{2}_{\rm \AdS{5}} + \sum_{i=1}^{3}\dd{\rho_{i}}^{2} + \rho_{i}^{2}\dd{\phi_{i}}^{2}\\
={}& \dd{s}^{2}_{\rm \AdS{5}} + \dd{\alpha}^{2} + \sin[2](\alpha) \dd{\theta}^{2} + \cos[2](\alpha) \pqty{\dd{\psi} - \dd{\varphi_{2}}}^{2} \\
&+ \sin[2](\alpha) \cos[2](\theta) \pqty{\dd{\psi} + \dd{\varphi_{1}}  + \dd{\varphi_{2}}}^{2} + \sin[2](\alpha) \sin[2](\theta) \pqty{ \dd{\psi} - \dd{\varphi_{1}}}^{2}\,.
\end{aligned}
\end{equation}
With the following identifications, we recover the metric~\eqref{eq:metric-S5}:
\begin{align}
\phi_{1} &= \psi -\varphi_{1}\,, & \phi_{2} &= \psi + \varphi_{1} + \varphi_{2}\,, & \phi_{3} &= \psi - \varphi_{2}.
\end{align}
These are the same angles that we have discussed in Sec.~\ref{sec:two-U1-fixed-points}:
\begin{align}
\varphi_{1} &= (-1,1,0)\,, & \varphi_{2} &= (0,1,-1)\,, & \psi &= (1,1,1)\,,
\end{align} 
which implies that the TsT transformation associated to $\psi$ does not preserve any supersymmetry.
If on the other hand we consider the TsT transformation \((\varphi_1, \varphi_2)_\lambda\) we obtain the following background:
\begin{equation}
\begin{aligned}
\dd{s}^{2} ={}& \dd{r}^{2} + e^{2r} \left( -(\dd{x^{0}})^{2} + (\dd{x^{1}})^{2} + (\dd{x^{2}})^{2} + (\dd{x^{3}})^{2}\right) + \dd{\alpha}^{2} + \sin[2](\alpha) \dd{\theta}^{2}\\
&+ \frac{1}{\Delta^{2}}\left(\sum_{i=1}^{3} \rho_{i}^{2} \dd{\phi_{i}}^{2} + \lambda^{2}\rho_{1}^{2}\rho_{2}^{2}\rho_{3}^{2}(\dd{\phi_{1}} + \dd{\phi_{2}} + \dd{\phi_{3}})^{2}\right)\,,\\
B_{2} ={}& -\frac{\lambda}{\Delta^{2}}(\rho_{1}^{2}\rho_{2}^{2}d\phi_{1}\wedge d\phi_{2} + \rho_{2}^{2}\rho_{3}^{2}d\phi_{2}\wedge d\phi_{3} + \rho_{3}^{2}\rho_{1}^{2}d\phi_{3}\wedge d\phi_{1})\,,\\
e^{-2\Phi} ={}& \Delta^{2}\,,\\
C_{2} ={}& \lambda \cos(\theta)\sin(\theta) \sin[4](\alpha) \dd{\theta} \wedge (\dd{\phi_{1}} + \dd{\phi_{2}} + \dd{\phi_{3}})\,,\\
C_{4} ={}& e^{4r} \dd{x^{0}} \wedge \dd{x^{1}} \wedge \dd{x^{2}} \wedge \dd{x^{3}} +\cos(\theta)\sin(\theta)\sin[4](\alpha) \dd{\theta} \wedge \dd{\phi_{1}} \wedge \dd{\phi_{2}} \wedge \dd{\phi_{3}}\,,\\
\Delta^{2} ={}& 1+ \lambda^{2}(\rho_{1}^{2}\rho_{2}^{2}+\rho_{2}^{2}\rho_{3}^{2}+\rho_{3}^{2}\rho_{1}^{2})\,,
\end{aligned}
\end{equation}
where $\rho_{1} = \cos \theta \sin \alpha\,, \rho_{2} = \sin\theta \sin \alpha\,, \rho_{3} = \cos\alpha$\,.
The bi-vector \(\Theta\) is \(\Theta = -\lambda \del_{\varphi_1} \wedge \del_{\varphi_2}\) and we would like to associate it to a non-commutativity of the ``quantum space'' type discussed in Section~\ref{sec:TsT-NC}.

As for the supersymmetry, the projectors related to $\varphi_{1}$ and $\varphi_{2}$, denoted by \(\proj{$\varphi_{1}$}\) and \(\proj{$\varphi_{2}$}\), are given in this case by 
\begin{equation}
\begin{aligned}
\proj{$\varphi_{1}$} &= \frac{1}{2} (1-\hat{\Gamma}_{\theta_{1}\phi_{2}\theta_{2}\phi_{1}})\,,\\
\proj{$\varphi_{2}$} & = \frac{1}{2}(1+\hat{\gamma}\hat{\Gamma}_{\phi_{3}\theta_{1}\phi_{2}})\,.
\end{aligned}
\end{equation} Their product preserves  one quarter of the supersymmetries in total.
The final result for the Killing spinors is
\begin{equation}
  \label{eq:Killing-after-ML}
  \begin{cases}
    \epsilon^{\text{(fin)}}_L = \epsilon^{\text{(in)}}_L \\
    \epsilon^{\text{(fin)}}_R = \frac{1}{\Delta}\biggl[1+\lambda (\rho_{1}\rho_{2}\hat{\Gamma}_{\phi_{1}\phi_{2}}+\rho_{2}\rho_{3}\hat{\Gamma}_{\phi_{2}\phi_{3}}+\rho_{3}\rho_{1}\hat{\Gamma}_{\phi_{3}\phi_{1}})\biggr]\epsilon^{\text{(in)}}_R ,
  \end{cases}
\end{equation}
where $\Delta = \sqrt{1+ \lambda^{2}(\rho_{1}^{2}\rho_{2}^{2}+\rho_{2}^{2}\rho_{3}^{2}+\rho_{3}^{2}\rho_{1}^{2})}$ and 
\begin{equation}
\begin{aligned}
\epsilon^{\text{(in)}} &= e^{\frac{i}{2}r\hat{\gamma}\Gamma_{r}}\left(1 + \frac{1}{2}x^{\alpha}\left( i \hat{\gamma}\hat{\Gamma}_{\alpha}+\hat{\Gamma}_{r}\hat{\Gamma}_{\alpha}\right)\right) \\
&\qquad \times e^{-\frac{1}{2}\theta_{2}\hat{\gamma}\hat{\Gamma}_{\theta_{2}}}e^{\frac{\theta_{1}}{2}\hat{\Gamma}_{\theta_{2}\theta_{1}}}e^{\frac{\psi}{2}(-\hat{\gamma}\hat{\Gamma}_{\phi_{3}}+\hat{\Gamma}_{\theta_{2}\phi_{1}}+\hat{\Gamma}_{\theta_{1}\phi_{2}})}\, \proj{$\varphi_{1}$}\proj{$\varphi_{2}$}\epsilon_{0}\,. 
\end{aligned}
\end{equation}
\subsubsection{A mixed example}

It is also possible to consider a TsT transformation involving isometries both in $\AdS{5}$ and $S^{5}$. Let us take the horospherical coordinates for $\AdS{5}$, given in~\eqref{eq:horos-coord}. Then the TsT $(x_{1}, \varphi_{1})_{\lambda}$ leads to the following background, where the \(C_4\) field is left untouched by the deformation:\footnote{As far as we know, this case has not yet been discussed in the context of the \ac{ybd}.}
\begin{equation}
\begin{aligned}
\dd{s}^2 ={}& \dd{r}^{2} + e^{2r}\left(-(\dd{x^{0}})^{2} +\frac{(\dd{x^{1}})^{2}}{\Delta^{2}}+ (\dd{x^{2}})^2 + (\dd{x^{3}})^{2} \right) + \dd{\alpha}^{2}+ \rho_{3}^{2} \left( \dd{\psi} - \dd{\varphi_{2}}\right)^{2}\\
& + \sin^{2}(\alpha) \dd{\theta}^{2} + \cos[2](\theta) \sin[2](\alpha) ( \dd{\psi} - \dd{\varphi_{1}})^{2} + \rho_{2}^{2} (\dd{\psi} + \dd{\varphi_{1}} + \dd{\varphi_{2}})^{2}\\
& + \frac{   1 - \Delta^{2} }{\Delta^{2}}\frac{\left(\rho_{1}^{2}(\dd{\varphi_{1}}-\dd{\psi})+\rho_{2}^{2}(\dd{\varphi_{1}+\dd{\varphi_{2}}+\dd{\psi}})\right)^{2}}{\rho_{1}^{2}+\rho_{2}^{2}}\,,\\
e^{-2\Phi} ={}& \Delta^{2}\\
B_{2} ={}& \frac{\Delta^{2} - 1}{ \lambda \Delta^{2}(\rho_{1}^{2}+\rho_{2}^{2})} \dd{x^{1}} \wedge ( \rho_{1}^{2}(\dd{\varphi_{1}}-\dd{\psi}) + \rho_{2}^{2}(\dd{\varphi_{1}} + \dd{\varphi_{2}} + \dd{\psi}))\,,\\
C_{4} ={}& e^{4r} \dd{x^{0}} \wedge \dd{x^{1}} \wedge \dd{x^{2}} \wedge \dd{x^{3}} - 3 \sin(\theta)\cos(\theta)\sin[4](\alpha) \dd{\theta} \wedge \dd{\varphi_{1}} \wedge \dd{\varphi_{2}} \wedge \dd{\psi}\,,
\end{aligned}
\end{equation}
where $\rho_{1} = \sin(\alpha)\cos(\theta), \rho_{2} = \sin(\alpha)\sin(\theta), \rho_{3} = \cos(\alpha)$, and $\Delta = \sqrt{1+\lambda^{2}e^{2r}(\rho_{1}^{2}+\rho_{2}^{2})}$.
The corresponding projector is given by
\begin{equation}
\proj{r} \proj{$\varphi_{1}$} = \frac{1}{2^{2}}(1+i\hat{\gamma}\hat{\Gamma}_{r})(1-\hat{\Gamma}_{\rho_{1}\phi_{1}\rho_{2}\phi_{2}})\,.
\end{equation}
The final result for the preserved Killing spinors is
\begin{equation}
  \label{eq:Killing-after-Mixed}
  \begin{cases}
    \epsilon^{\text{(fin)}}_L = \epsilon^{\text{(in)}}_L \\
    \epsilon^{\text{(fin)}}_R = \frac{1}{\Delta}\biggl[1+\lambda e^{r} (\rho_{2}\hat{\Gamma}_{x^{1}\phi_{1}}-\rho_{1}\hat{\Gamma}_{x^{1}\phi_{2}})\biggr]\epsilon^{\text{(in)}}_R ,
  \end{cases}
\end{equation}
with 
\begin{equation}
\begin{aligned}
\epsilon^{\text{(in)}} &= e^{\frac{i}{2}r\hat{\gamma}\Gamma_{r}}e^{-\frac{1}{2}\theta_{2}\hat{\gamma}\hat{\Gamma}_{\theta_{2}}}e^{\frac{\theta_{1}}{2}\hat{\Gamma}_{\theta_{2}\theta_{1}}}e^{\frac{\psi}{2}(-\hat{\gamma}\hat{\Gamma}_{\phi_{3}}+\hat{\Gamma}_{\theta_{2}\phi_{1}}+\hat{\Gamma}_{\theta_{1}\phi_{2}})}e^{\frac{\varphi_{2}}{2}(\hat{\gamma}\hat{\Gamma}_{\phi_{3}}+\hat{\Gamma}_{\theta_{1}\phi_{2}})}\, \proj{$r$}\proj{$\varphi_{1}$}\epsilon_{0}\,. 
\end{aligned}
\end{equation}
\subsection{Schr\"odinger spacetime: a non-supersymmetric example}

Many examples of integrable deformations break all supersymmetries. For completeness, 
we will therefore also discuss such an example, namely Schr\"odinger spacetime~\cite{arXiv:0807.1099,arXiv:0807.1100,arXiv:0807.1111}\footnote{The three-parameter generalization is discussed in~\cite{arXiv:0905.0673}.}.  
This is again a deformation of $\AdS{5} \times S^5$ with mixed type.
One can write $S^{5}$ as an $S^{1}$ fibration over $\mathbb{C}\mathrm{P}^{2}$. The metric is characterized by the $S^{1}$-fiber $\chi$: 
\begin{equation}
\begin{aligned}
\dd{s}^{2}_{\rm S^{5}} &= (\dd{\chi} + \omega)^{2} + \dd{s}^{2}_{\mathbb{C}\mathrm{P}^{2}}\,,\\
\dd{s}^{2}_{\mathbb{C}\mathrm{P}^{2}} &= \dd{\mu}^{2} + \sin[2]( \mu) \left( \Sigma^{2}_{1} + \Sigma^{2}_{2} + \cos[2](\mu) \Sigma_{3}^{2}\right)\,,
\end{aligned}
\end{equation}
where $\Sigma_{i}$ for $i=1,2,3$ and $\omega$ are given by
\begin{equation}
\begin{aligned}
\Sigma_{1} &= \frac{1}{2}(\cos(\psi) \dd{\theta} + \sin(\psi) \sin(\theta) \dd{\tau} )\,,\\
\Sigma_{2} &= \frac{1}{2}(\sin(\psi) \dd{\theta} - \cos(\psi) \sin(\theta) \dd{\tau} )\,,\\
\Sigma_{3} &= \frac{1}{2}(\dd{\psi} + \cos( \theta) \dd{\tau} )\,,\\
\omega &= \sin^{2}\mu \Sigma_{3}\,.
\end{aligned}
\end{equation}
The coordinates $(\chi, \psi, \tau)$ are related to $(\phi_{1}, \phi_{2}, \phi_{3})$ by\footnote{For the other two coordinates, $\theta_{2} = \mu$ and $\theta_{1} = \frac{1}{2}\theta$. } 
\begin{equation}
\begin{aligned}
\phi_{1} &= \chi + \frac{1}{2}(\psi + \tau)\,,\\
\phi_{2} &= \chi + \frac{1}{2}(\psi - \tau)\,,\\
\phi_{3} &= \chi\,. 
\end{aligned}
\end{equation}
In our notation,
\begin{align}
\chi &= (1,1,1)\,, & \psi &= \frac{1}{2}(1,1,0)\,, & \tau &= \frac{1}{2}(1,-1,0)\,.
\end{align}
Schr\"odinger spacetime is obtained as the TsT transformation $(\chi, x^{-})_{\lambda}$, where $x^- =x^0 - x^1$ lives in $\AdS{5}$. 
Supersymmetry is completely broken by $(\chi, x^{-})_{\lambda}$, since $\chi$ does not satisfy any of the conditions listed in Table~\ref{tab:SUSY}.
The TsT-transformations $(\psi, x^{-})_{\lambda}$ and $(\tau, x^{-})_{\lambda}$ preserve $\frac{1}{4}$-supersymmetry. 

This TsT transformation can be understood as a \ac{ybd}~\cite{arXiv:1502.00740}.

\subsection{Killing spinors for deformations of 11d flat space}

Although there is no T-duality in M-theory itself, we can perform the analogue of a TsT transformation also in M-theory as discussed in Section~\ref{sec:TsT-M}. 

As an illustrative example, we start from eleven-dimensional flat space, where two $ \setR ^{2}$ planes are expressed in polar coordinates,
\begin{equation}
\dd{s}^{2} = -(\dd{x^{0}})^{2} + \sum_{m=1}^{4} (\dd{x^{m}})^{2} + \dd{u}^{2} + \sum_{i=1}^{2} \dd{\rho_{i}}^{2} + \rho_{i}^{2} \dd{\phi_{i}}^{2}+ (\dd{x^{10}})^{2} \,,
\end{equation}
where $u$ and $x^{10}$ are periodic directions. There is only one choice of $T^3$ which preserves supersymmetry, namely the transformation $(u, \varphi, x^{10})_\lambda$, with $\varphi = \tfrac{1}{2}(\phi_1-\phi_2)$. Since there is no four-form flux, the complex parameter $\tau_{M}$ in~\eqref{eq:tauM} is always purely imaginary. %

This deformation leads to the so-called M-theory fluxtrap background~\cite{arXiv:1204.4192}:
\begin{equation}
\begin{aligned}
\dd{s}^{2} ={}& \Delta^{2/3}\left(-(\dd{x^{0}})^{2} + \sum_{m=1}^{4} (\dd{x^{m}})^{2} + \sum_{i=1}^{2} \dd{\rho_{i}}^{2} + \rho_{i}^{2} \dd{\phi_{i}}^{2}\right) \\
&  - \lambda^{2}\Delta^{-4/3}\left(\rho_{1}^{2} \dd{\phi_{1}} - \rho_{2}^{2} \dd{\phi_{2}}\right)^{2} + \Delta^{-4/3}\left(\dd{u}^{2} + (\dd{x^{10}})^{2}\right),\\
C_{3} &= -\lambda \frac{\rho_{1}^{2} \dd{\phi_{1}} - \rho_{2}^{2} \dd{\phi_{2}}}{\Delta^{2}}\wedge \dd{u} \wedge \dd{x^{10}},\\
\tau_{M}^{\prime} &= \frac{i \sqrt{\rho_{1}^{2}+\rho_{2}^{2}}}{1+i \lambda \sqrt{\rho_{1}^{2}+\rho_{2}^{2}}}\,,
\end{aligned}
\end{equation}
where $\Delta = \sqrt{1 + \lambda^{2}(\rho_{1}^{2}+\rho_{2}^{2})} = \abs{1+ \lambda \tau_{M}}$. 
Just like in the ten-dimensional case, the preserved Killing spinors are those which do not depend on any of the directions generating the $T^3$. In this case, they are projected out by  
\begin{equation}
\proj{$\varphi$} = \frac{1}{2} \pqty{1 - \hat{\Gamma}_{\rho_1 \phi_1 \rho_2 \phi_2}} .
\end{equation}
and the preserved spinors in the initial frame are expressed by
\begin{equation}
\epsilon^{\text{(in)}}_{M} = \exp{\frac{\phi_{1} + \phi_{2}}{4}(\hat{\Gamma}_{\rho_{1}\phi_{1}}-\hat{\Gamma}_{\rho_{2}\phi_{2}})}\proj{$\varphi$}\epsilon_{0}.
\end{equation}
Note that there is a relation between Killing spinors in type IIA and M-theory as follows:
\begin{equation}
\epsilon_{M} = e^{-\Phi/6}\epsilon_{IIA}\,.
\end{equation}
Therefore, via M-theory TsT, we find the following rule
\begin{equation}
\epsilon_{M}^{(2)} = e^{(\Phi^{(1)}-\Phi^{(2)})/6}\biggl[P_{L} + \Gamma^{{\text{fin}}}_{u}\Gamma^{{\text{in}}}_{u}P_{R}\biggr]\epsilon_{M}^{(1)}\,,
\end{equation}
where $P_{L/R}$ are projectors defined in type IIA. Then the final result for the Killing spinor is given by
\begin{equation}
\epsilon^{\text{(fin)}}_{M} = \Delta^{1/6}\biggl[P_{L} + \frac{1}{\Delta}(1-\lambda(\rho_{1}\hat{\Gamma}_{u\phi_{1}}-\rho_{2}\hat{\Gamma}_{u\phi_{2}}))P_{R}\biggr]\epsilon_{M}^{\text{(in)}}\,.
\end{equation}
It had already been observed~\cite{arXiv:1204.4192,Lambert:2013lxa,arXiv:1409.1219} that $u$ and $x^{10}$ appear completely symmetrically in the background. From the string theory point of view, the reason for this is unclear, but here we can understand it as an effect of the $SL(3)$ action which is geometrical in M-theory, but not in string theory.

\subsection{Killing spinors for deformations of $\AdS{7} \times S^4$}

In Section~\ref{sec:AdS5}, we discussed the recipe for obtaining the Killing spinor for deformations of $\AdS{5} \times S^5$, which can be easily generalized to the case of   $\AdS{7} \times S^4$.  Again, the condition for preserving any supersymmetry is that the Killing spinors do not depend on \emph{any} of the directions in the $T^3$. 

We consider one particular choice of frame which is particularly illustrative as it contains  $U(1)$s that act both freely and non-freely in the embedding space. 

We consider the following metric for $\AdS{7}\times S^{4}$:
\begin{multline}
  \dd{s_{\AdS{7}\times S^{4}}}^{2} = \dd{r}^{2} + e^{r} \pqty{ -(\dd{x^0})^{2} + (\dd{x^{1}})^{2} + \sum_{i=1}^{2}\pqty{\dd{\rho_{i}}^{2} + \rho_{i}^{2} \dd{\varphi_{i}}}} \\ + \dd{\theta_{2}}^{2} + \sin[2](\theta_2)  \pqty{\dd{\theta_{1}}^{2} + \cos[2](\theta_1) \dd{\phi_{1}}^{2} + \sin[2](\theta_1) \dd{\phi_{2}}^{2}}\,.
\end{multline}
For the three-form, we pick a gauge such that it is invariant under the $U(1)^{2}$ acting on $\phi_{1}$ and $\phi_{2}$:
\begin{equation}
  C_{3} = -\frac{3}{4}\cos(2\theta_{1}) \sin[3](\theta_2) \dd{\theta_{2}} \wedge \dd{\phi_{1}} \wedge \dd{\phi_{2}}\,.
\end{equation}
Note that the complex parameter \(\tau_M\) is always purely imaginary for all possible choices of \(T^3\) because \(C_3\) always has one leg in the direction \(\theta_2\) that is not an isometry.

\bigskip

As shown in Appendix~\ref{app:KillingAdS7}, the Killing spinor obeys almost the same equations as in $\AdS{5}\times S^{5}$, and is given by
\begin{equation}
\begin{aligned}
\epsilon_{\AdS{7}\times S^{4}} ={}&  e^{\frac{1}{4}r\hat{\gamma}\hat{\Gamma}_{r}}\biggl[1+\frac{1}{4}\sum_{\substack{\alpha=0,1\\
    i=1,2}}\left(x^{\alpha}\left(  \hat{\gamma}\hat{\Gamma}_{\alpha}+\hat{\Gamma}_{r}\hat{\Gamma}_{\alpha}\right)+\rho_{i}\left(  \hat{\gamma}\hat{\Gamma}_{\rho_{i}}+\hat{\Gamma}_{r}\hat{\Gamma}_{\rho_{i}}\right)\right)\biggr] \\
& \times e^{\frac{1}{2}\varphi_{1}\hat{\Gamma}_{\rho_{1}\varphi_{1}}}e^{\frac{1}{2}\varphi_{2}\hat{\Gamma}_{\rho_{2}\varphi_{2}}}e^{\frac{1}{2}\theta_{2}\hat{\gamma} \hat{\Gamma}_{\theta_{2}}}e^{-\frac{1}{2}\theta_{1}\hat{\Gamma}_{\theta_{1}\theta_{2}}}e^{\frac{1}{2}\phi_{1}\hat{\Gamma}_{\theta_{2}\phi_{1}}}e^{\frac{1}{2}\phi_{2}\hat{\Gamma}_{\theta_{1}\phi_{2}}}\epsilon_{0}\,,
\end{aligned}
\end{equation}
where $\hat{\gamma} = \hat{\Gamma}_{89(10)(11)}$ in our conventions. We now present two examples illustrating the different possible choices of the $T^3$ in $\AdS{7} \times S^4$.
We define the following angles:
\begin{equation}
\begin{aligned}
\varphi_{\pm} = \frac{\varphi_{1}\pm \varphi_{2}}{2}\,,\qquad \phi_{\pm} = \frac{\phi_{1}\pm\phi_{2}}{2}\,.
\end{aligned}
\end{equation}

\subsubsection{$(\phi_{+}, x^{1}, \varphi_{+})_{\lambda}$-type}

To begin with, we choose the coordinate $x_{1}$ to be periodic, imposing an additional boundary condition on $\AdS{7}$. 
 Starting from the the three-torus generated by $(\phi_{+}, x_{1}, \varphi_{+})$ with the parameter
\begin{equation}
\tau_{M} = i e^{r}\abs{\sin(\theta_2)} \sqrt{\rho_{1}^{2}+\rho_{2}^{2}}\,,
\end{equation}
we obtain the following background:
\begin{equation}
  \begin{small}
\begin{aligned}
\dd{s}^{2} ={}& \Delta^{2/3}\left(\dd{r}^{2} + e^{r}\left(-(\dd{x^{0}})^{2} + (\dd{x^{1}})^{2} + \sum_{i=1}^{2} \dd{\rho_{i}}^{2} + \rho_{i}^{2} \dd{\varphi_{i}}^{2} + \frac{(\rho_{i}^{2} \dd{\varphi_{i}})^{2}}{\rho_{1}^{2} + \rho_{2}^{2}}\right) + \sin(\theta_2)^{2}\sin[2](2\theta_{1}) \dd{\phi_{-}}^{2}\right)\\
& + \Delta^{-4/3}\left(e^{r}\left( (\dd{x^{1}})^{2} -\frac{(\rho_{1}^{2} \dd{\varphi_{1}} + \rho_{2}^{2} \dd{\varphi_{2}})^{2}}{\rho_{1}^{2} + \rho_{2}^{2}}\right) + \sin(\theta_2)^{2}( \dd{\phi_{+}} + \cos(2\theta_{1}) \dd{\phi_{-}})^{2}\right)\,,\\
C_{3} ={}& -\frac{3}{4}\sin(\theta_2)^{3}\cos(2\theta_{1}) \dd{\theta_{2}} \wedge \dd{\phi_{1}}\wedge \dd {\phi_{2}}\\
& + \frac{1-\Delta^{2}}{\lambda \Delta^{2}} \dd{x^{1}}\wedge \left(\cos(2\theta_{1}) \dd{\phi_{-}} \wedge \dd{\varphi_{+}} + 2\frac{\rho_{1}^{2} - \rho_{2}^{2}}{\rho_{1}^{2}+\rho_{2}^{2}}(\dd{\phi_{+}} + 3\cos(2\theta_{1})\dd{\phi_{-}}) \wedge \dd{\varphi_{-}}\right)\\
& + \frac{\Delta^{2}-1}{\lambda \Delta^{2}} \dd{\phi_{+}} \wedge \dd{x_{1}} \wedge \dd{\varphi_{+}}\,,\\
\tau_{M}^{\prime} ={}& C_{\phi_{+}x^{1}\varphi_{+}} + i \sqrt{g_{(\phi_{+}x^{1}\varphi_{+})}}\\
={}& \frac{1}{\lambda \Delta^{2}}\left(\Delta^{2}-1 + i \sqrt{\Delta^{2}-1}\right) = \frac{\tau_{M}}{1 + \lambda \tau_{M}}\,,
\end{aligned}
\end{small}
\end{equation}
with $\Delta = \sqrt{1 + \lambda^{2} e^{2r} \sin(\theta_2)^{2}(\rho_{1}^{2}+\rho_{2}^{2})}= \abs{1+\lambda \tau_M}$\,.

The supersymmetries preserved under this M-theory TsT transformation are those which are not projected out by
\begin{equation}
\proj{MTsT} = \proj{$\phi_{+}$}\proj{r}\proj{$\varphi_{+}$} = \frac{1}{2^{3}}(1+\hat{\Gamma}_{\theta_{2}\phi_{1}\theta_{1}\phi_{2}})(1+\hat{\gamma}\hat{\Gamma}_{r})(1+\hat{\Gamma}_{\rho_{1}\varphi_{1}\rho_{2}\varphi_{2}})\,,
\end{equation}
which means that there are only four surviving Killing spinors.
The preserved components of the initial Killing spinor are given by
\begin{equation}
\begin{aligned}
\epsilon_{M}^{\text{(in)}} ={}&  e^{\frac{1}{4}r\hat{\gamma}\hat{\Gamma}_{r}} e^{\varphi_{-}\hat{\Gamma}_{\rho_{1}\varphi_{1}}}e^{\frac{1}{2}\theta_{2}\hat{\gamma} \hat{\Gamma}_{\theta_{2}}}e^{-\frac{1}{2}\theta_{1}\hat{\Gamma}_{\theta_{1}\theta_{2}}}e^{\phi_{-}\hat{\Gamma}_{\theta_{2}\phi_{1}}}\proj{MTsT}\epsilon_{0}\,.
\end{aligned}
\end{equation}
Also, the Killing spinor after the deformation is of the form
\begin{equation}
\epsilon^{\text{(fin)}}_{M} = \Delta^{1/6}\biggl[P_{L} + \frac{1- \lambda e^{r}\sin\theta_{2}\sqrt{\rho_{1}^{2}+\rho_{2}^{2}}(\cos\theta_{1}\hat{\Gamma}_{\phi_{1}x_{1}}+\sin\theta_{1}\hat{\Gamma}_{\phi_{2}x_{1}})}{\Delta}P_{R}\biggr]\epsilon^{\text{(in)}}_{M}\,.
\end{equation}
\subsubsection{Euclideanized $(x^{0}, \varphi_{-}, x^{1})$-type}

To be able to perform a supersymmetric TsT transformation only within $\AdS{7}$, we need to analytically continue in $x^{0}$. Then the three-torus is generated by $(x^{0}, \varphi_{-}, x^{1})$ with
\begin{equation}
\tau_{M} = i e^{3r/2}\sqrt{\rho_{1}^{2}+\rho_{2}^{2}}. 
\end{equation}
The deformed background is expressed with $\Delta = \sqrt{1 + \lambda^{2}e^{3r}(\rho_{1}^{2}+\rho_{2}^{2})} = \abs{1+\lambda \tau_M}$ as
\begin{equation}
\begin{aligned}
\dd{s}^{2} ={}& \Delta^{2/3}\left( \dd{r}^{2} + e^{r}( \dd{ \rho_{1}}^{2} + \dd{\rho_{2}}^{2} + \frac{\rho_{1}^{2}\rho_{2}^{2}(\dd{\varphi_{1}}+\dd{\varphi_{2}})^{2}}{\rho_{1}^{2}+\rho_{2}^{2}}) + \dd{s_{S^{4}}}^{2}\right)\\
& + \Delta^{-4/3}e^{r}\left( (\dd{x^{0}})^{2} + (\dd{x^{1}})^{2} + \frac{(\rho_{1}^{2} \dd{\varphi_{1}}-\rho_{2}^{2} \dd{\varphi_{2}})^{2}}{\rho_{1}^{2}+\rho_{2}^{2}} \right) ,\\
C_{3} ={}& -\frac{3}{4}s_{2}^{3}\cos(2\theta_{1}) \dd{\theta_{1}} \wedge \dd{\phi_{1}} \wedge \dd{\phi_{2}} - \frac{\Delta^{2}-1}{\lambda \Delta^{2}}\left(\frac{\rho_{1}^{2}-\rho_{2}^{2}}{\rho_{1}^{2}+\rho_{2}^{2}} \dd{\varphi_{+}} + \dd{\varphi_{-}}\right) \wedge \dd{x^{0}} \wedge \dd{x^{1}} , \\
\tau^{\prime}_{M} ={}& \frac{\Delta^{2}-1}{\lambda \Delta^{2}} + i \frac{\sqrt{\Delta^{2}-1}}{\Delta^{2}} = \frac{\tau_{M}}{1+ \lambda \tau_{M}}\,.
\end{aligned}
\end{equation}
This M-theory TsT transformation preserves eight Killing spinors surviving the projector 
\begin{equation}
\proj{MTsT} = \proj{r} \proj{$\varphi_{-}$} = \frac{1}{2^{2}}(1+\hat{\gamma}\hat{\Gamma}_{r})(1-\hat{\Gamma}_{\rho_{1}\varphi_{1}\rho_{2}\varphi_{2}})\,.
\end{equation}
Out of the initial 32-components, the following are compatible with the M-theory TsT transformation:
\begin{equation}
\begin{aligned}
\epsilon_{M}^{\text{(in)}} ={}&  e^{\frac{1}{4}r\hat{\gamma}\hat{\Gamma}_{r}}e^{\varphi_{+}\hat{\Gamma}_{\rho_{1}\varphi_{1}}}e^{\frac{1}{2}\theta_{2}\hat{\gamma} \hat{\Gamma}_{\theta_{2}}}e^{-\frac{1}{2}\theta_{1}\hat{\Gamma}_{\theta_{1}\theta_{2}}}e^{\frac{1}{2}\phi_{1}\hat{\Gamma}_{\theta_{2}\phi_{1}}}e^{\frac{1}{2}\phi_{2}\hat{\Gamma}_{\theta_{1}\phi_{2}}}\proj{MTsT}\epsilon_{0}\,.
\end{aligned}
\end{equation}
Finally, the Killing spinor transforms as
\begin{equation}
\epsilon_{M}^{\text{(fin)}} = \Delta^{1/6}\biggl[P_{L} + \frac{1-\lambda e^{3r/2}(\rho_{1}\hat{\Gamma}_{x^{0}\varphi_{1}}-\rho_{2}\hat{\Gamma}_{x^{0}\varphi_{2}})}{\Delta}P_{R}\biggr]\epsilon^{\text{(in)}}_{M}\,.
\end{equation}

\section{Conclusions}\label{sec:conclusions}

In this article, we have discussed the interplay between integrability, supersymmetry and non-commutativity for \acp{ybd} with unimodular classical $r$-matrices 
satisfying the homogeneous \ac{cybe}.
Some deformations in this class can be realized in type~II string theory compactified on \(T^2\) as TsT transformations that correspond to the transformation \(\tau \to \tau / ( 1 + \lambda \tau )\) of the K\"ahler parameter \(\tau = B_{12} + i \sqrt{g_{(12)}}\).
In flat space we identify three families of TsT transformations, depending on the action (free or non-free) of the two \(U(1)\) symmetries in the torus directions.
We find a similar classification for the case of \(\AdS{5} \times S^5\) as well.
These three families are naturally associated to three types of non-commutative deformations of gauge theories in the classification of~\cite{Wess:2003da,Lukierski:2005fc}, namely the canonical, Lie algebra and quantum space cases.

All the data of the deformation can be encoded in a ten-dimensional bi-vector \(\Theta\), which in the context of \textsc{yb}-deformed integrable models has a natural interpretation in terms of classical $r$-matrices and in the context of gauge theory as a non-commutativity parameter.
Our main result is an explicit construction of the Killing spinors of the TsT-deformed theory, which is again completely encoded in the bi-vector \(\Theta\). 
We decompose \(\Theta \) into components and associate projectors on the space of ten-dimensional spinors  to them.
In this way we can also write explicit expressions for the Killing spinors in adapted frames, which we discuss in detail for the cases of \(\AdS{5}\) and \(S^5\):
\begin{align}
  \epsilon_{L}^{\text{(fin)}} &= \proj{TsT} \epsilon_L^{\text{(in)}} , &
  \epsilon_{R}^{\text{(fin)}} &=  e^{\omega(\Theta) \, \Theta^{\mu\nu}\hat{\Gamma}_{\mu}\hat{\Gamma}_{\nu}}\proj{TsT} \epsilon^{\text{(in)}}_{R}\,, \label{formula}
\end{align}
where \(\epsilon^{\text{(in)}}\) are the Killing spinors in the initial background, \(\omega(\Theta)\) is normalization factor which is a function of \(\Theta^{\mu \nu } \Theta_{\mu \nu}\), and \(\proj{TsT}\) is an appropriate projector that is completely determined by \(\Theta\).
{The exponential factor in the formula (\ref{formula}) is very suggestive 
since this form may indicate a relationship to the quantum $R$-matrix in the spinorial  
representation. We anticipate that this factor is related to the phase factor 
for the fermions discussed in~\cite{Beisert:2005if} in some manner.}

The conditions for the preservation of supersymmetry can then be naturally recast in the language of the \(\varepsilon\) parameters of the \(\Omega\)-deformation.
The key observation is that the fluxtrap background used to realize the \(\Omega\)-deformation in string and M-theory is a special case of the TsT transformations that we discuss here (namely the one corresponding to Lie-algebra-type non-commutativity).
This observation allows us to relate the \(\Omega\)-deformation directly to integrability on the one hand and to non-commutativity on the other and opens a number of directions of research that deserve further investigation.

The TsT construction of type~II string theory has a natural generalization to M-theory compactified on \(T^3\).
We discuss this generalization and we give a recipe for the explicit construction of the corresponding Killing spinors in eleven dimensions.
From the point of view of supergravity, the generalization to eleven dimensions is straightforward and natural.
It remains an open question whether this transformation has an interpretation in terms of integrability and/or non-commutativity.
We intend to revisit these issues in the future.

\medskip
This work opens up also other natural directions of investigation.
We enumerate here a few of them.
\begin{itemize}
\item We have seen that the \(\Omega\)-deformation has a natural interpretation as a TsT deformation.
  Can this lead us directly to the realization of its gravity dual?
\item We have seen that the TsT transformations are naturally associated to classical $r$-matrices and to constant, linear and quadratic non-commutativity in gauge theory.
  However, in this latter context it is possible to have more general \ac{nc} parameters.
  Do they have an interpretation in terms of $r$-matrices and/or transformations in string and M-theory?
\item From the string theory perspective only one family of TsT transformations (the one corresponding to the canonical \ac{nc} parameter) has a precise interpretation in terms of non-commutativity via the Seiberg--Witten map~\cite{hep-th/9908142}.
  On the other hand, in all the cases we have studied we can at least formally apply the same map.
  Is it possible to consistently generalize the work of Seiberg and Witten to the other two families of deformations?
\item We have a natural lift to M-theory compactified on a three-torus.
  This seems to be a natural setup to study S-duality.
  Can we say anything about the relationship between integrability and S-duality from this point of view?
\end{itemize}

\subsection*{Acknowledgments}

{It is our pleasure to thank J.~Sakamoto and Y.~Sakatani for useful discussion and comments on the draft.}
D.O. would like to thank C.~Angelantonj for discussions.
D.O. and S.R. would like to thank the Department of Physics of Kyoto University for hospitality.  D.O. and S.R. would also like to thank the Galileo Galilei Institute for Theoretical Physics for hospitality and the \textsc{infn} for partial support during the completion of this work. 
The work of S.R. and Y.S. is supported by the Swiss National Science Foundation (\textsc{snf}) under grant number PP00P2\_157571/1.
D.O. acknowledges partial support by the NCCR 51NF40-141869 ``The Mathematics of Physics'' (SwissMAP). 
The work of K.Y. is supported by the Supporting Program for Interaction based Initiative Team Studies 
(SPIRITS) from Kyoto University and by a JSPS Grant-in-Aid for Scientific Research (B) No.\,18H01214  
and was supported by a JSPS Grant-in-Aid for Scientific Research (C) No.\,15K05051. 
This work is also supported in part by the JSPS Japan-Russia Research Cooperative Program.

\appendix
\section{Gamma matrices}\label{app:Gamma}
We parametrize the Gamma matrices based on the conventions in~\cite{Lu:1998nu}. The five-dimensional Gamma matrices are defined as
\begin{equation}
\label{eq:Gamma-matrix}
\begin{aligned}
\Gamma_{1} &= \sigma_{1} \otimes \mathbf{1} \otimes \mathbf{1} \otimes\cdot\cdot\cdot \otimes\mathbf{1}\,,\quad &\Gamma_{2} &= \sigma_{2} \otimes \mathbf{1} \otimes \mathbf{1} \otimes\cdot\cdot\cdot \otimes \mathbf{1}\,,\\
\Gamma_{3} &= \sigma_{3} \otimes \sigma_{1} \otimes \mathbf{1} \otimes \cdot\cdot\cdot \otimes\mathbf{1}\,,\quad &\Gamma_{4} &= \sigma_{3} \otimes \sigma_{2} \otimes \mathbf{1} \otimes \cdot\cdot\cdot \otimes \mathbf{1}\,,\\
\cdot\cdot\cdot\\
\Gamma_{2n-1} &= \sigma_{3} \otimes \sigma_{3} \otimes \cdot\cdot\cdot \otimes \sigma_{3} \otimes \sigma_{1}\,,\quad & \Gamma_{2n} &= \sigma_{3} \otimes \sigma_{3} \otimes \cdot\cdot\cdot \otimes \sigma_{3} \otimes \sigma_{2},
\end{aligned}
\end{equation}
where $\sigma_i~(i=1,2,3)$ are the standard Pauli matrices. 
The imaginary $i$ appears in front when considering the metric with the Lorentzian signature.
It is convenient to decompose the ten-dimensional Dirac matrices into lower-dimensional ones and to use the latter in the calculations. In the case of $\AdS{5} \times S^{5}$\,, we decompose the ten-dimensional Gamma matrices, denoted by $\hat{\Gamma}_{A}$, as follows:
\begin{equation}
\label{eq:gamma-decomposition}
\begin{aligned}
\AdS{5}:& \quad \ \hat{\Gamma}_{\mu} =\,  \sigma_{1} \otimes \Gamma_{\mu} \otimes \mathbf{1}_{4\times4}\,,\\
S^{5}: & \quad \ \hat{\Gamma}_{m} =\, \sigma_{2} \otimes \mathbf{1}_{4\times4} \otimes \Gamma_{m}\,,
\end{aligned}
\end{equation}
where {the $\sigma_{1,2}$ are needed} to satisfy the Clifford Algebra in ten dimensions. 
Using this parametrization, the ten-dimensional chirality matrix is explicitly written as
\begin{equation}
\begin{aligned}
\hat{\Gamma}_{(10)} &= \hat{\Gamma}_{0} \cdot\cdot\cdot \hat{\Gamma}_{9}\\
&=\sigma_{3} \otimes \mathbf{1}_{4\times4} \otimes \mathbf{1}_{4\times4}\,,
\end{aligned}
\end{equation}
which automatically gives a positive chirality for the Killing spinor of the form
\begin{equation}
\label{eq:spinor-decomposition}
\epsilon =
\begin{pmatrix}
1\\0
\end{pmatrix}
\otimes \epsilon_{\AdS{5}}
\otimes \epsilon_{S^{5}} .
\end{equation}
On the other hand, in $\AdS{7} \times S^{4}$, the Gamma matrices can be decomposed without Pauli matrices as follows:
\begin{equation}\label{eq:Gamma-Ads7}
\begin{aligned}
\AdS{7}:& \quad \ \hat{\Gamma}_{\mu} =\,  \Gamma_{\mu} \otimes \gamma\,,\\
S^{4}: & \quad \ \hat{\Gamma}_{j} =\, \mathbf{1}_{8\times8} \otimes \Gamma_{j}\,,
\end{aligned}
\end{equation}
where $\gamma = \Gamma_{89(10)(11)}$ is a chirality operator satisfying $\gamma^{2} = 1$. 
Likewise, the eleven-dimensional Killing spinor decomposes into two parts:
\begin{equation}\label{eq:spinor-Ads7}
\epsilon = \epsilon_{\AdS{7}} \otimes \epsilon_{S^{4}}\,.
\end{equation}
Note that the chirality operator is uplifted to 
\begin{equation}
\hat{\gamma} = \mathbf{1}_{8\times8}\otimes \gamma
\end{equation}
in the eleven-dimensional expression.

\section{Killing spinors for undeformed $\AdS{5} \times S^{5}$}
\label{sec:spinor-detail}
The preceding appendix enables us to solve the gravitino equation for the $\AdS{5}\times S^{5}$ background,
\begin{equation}
\label{eq:gravitino-eq}
D_{A}\epsilon + \frac{i}{2^{4}\cdot5!}F_{MNPQR}\hat{\Gamma}^{MNPQR}\hat{\Gamma}_{A}\epsilon = 0\,
\end{equation}
 in terms of two sectors. Using the decomposition of the Gamma matrices~\eqref{eq:gamma-decomposition} as well as Killing spinors~\eqref{eq:spinor-decomposition}, we find 
 \begin{equation}
 \begin{aligned}
 \AdS{5}: \quad \ &D_{\mu}\epsilon_{\AdS{5}} = \frac{1}{2}\Gamma_{\mu}\epsilon_{\AdS{5}},\\
 S^{5}: \quad \ &D_{j} \epsilon_{S^{5}} = \frac{i}{2}\Gamma_{j}\epsilon_{S^{5}}.
 \end{aligned}
 \end{equation}
 We now present the computational details for the Killing spinor on an ordinary $\AdS{5}\times S^{5}$.

\subsection{$\AdS{5}$}
Let us express the metric in the horospherical coordinates:
\begin{equation}\label{eq:AdS5HS}
\dd{s}^{2}_{\AdS{5}}= e^{2r}\eta_{\alpha\beta} \dd{x^{\alpha}} \dd{x^{\beta}} + \dd{r^{2}}\,,
\end{equation}
where $\eta_{\alpha\beta} = \mathrm{diag}(-,+,+,+)$. A set of the vielbein one-forms is naturally defined as
\begin{align}
\mathbf{e}^{\alpha} &= e^{r} \dd{x^{\alpha}}\,, \quad \mathbf{e}^{r} = \dd{r}\,.
\end{align}
Hence, the Maurer--Cartan equation leads to the spin connections given by
\begin{equation}
\begin{aligned}
\omega^{\alpha r}_{\alpha} = e^{r}\,,\quad \omega^{\alpha r}_{r} = 0\,,
\end{aligned}
\end{equation}
and the gravitino equation~\eqref{eq:gravitino-eq} reads
\begin{equation}
\begin{aligned}
\partial_{\alpha}\epsilon &= \frac{e^r}{2}\Gamma_{\alpha}\left(1 - \Gamma_{r}\right)\epsilon\,,\\
\partial_{r}\epsilon &=\frac{1}{2}\Gamma_{r}\epsilon\,.
\end{aligned}
\end{equation}
The first equation allows us to solve the equations using the projector for the $r$-direction. Consequently, the collective solution is of the form
\begin{equation}
\label{eq:spinor-ads5}
\epsilon_{\AdS{5}} = e^{\frac{1}{2}r\Gamma_{r}}\left(1 + \frac{1}{2}x^{\alpha}\Gamma_{\alpha}(1 - \Gamma_{r})\right)\epsilon_{0}\,,
\end{equation}
where $\epsilon_{0}$ is a four-component constant spinor, and the Gamma matrices $\Gamma_{\alpha}$ in the parent have flat indices and correspond to~\eqref{eq:Gamma-matrix} as follows:
\begin{equation}
\Gamma_{x^{0}} = i \Gamma_{1}\,,\quad \Gamma_{x^{1}} = \Gamma_{2}\,,\quad \Gamma_{x^{2}} = \Gamma_{3}\,,\quad \Gamma_{x^{3}} = \Gamma_{4}\,,\quad \Gamma_{r} = \Gamma_{5}. 
\end{equation}

\subsection{$S^{5}$}
The five-sphere is parametrized such that the $U(1)^{3}$ symmetry is manifest. 
The metric is given by
\begin{equation}
  \begin{aligned}
  \dd{s}_{S^{5}}^{2} &= \sum_{i=1}^{3}\dd{\rho_{i}}^{2} + \rho_{2}^{2} \dd{\phi_{i}}^{2}\\
   &=
   \begin{multlined}[t]
\dd{\theta_2}^{2} + \sin[2](\theta_2) \dd{\theta_1}^{2}  + \sin[2](\theta_2)  \cos[2](\theta_1) \dd{\phi_1}^{2} \\
  + \sin[2](\theta_2) \sin[2](\theta_1) \dd{\phi_2}^{2} + \cos[2]( \theta_2) \dd{\phi_3}^{2} \,, 
\end{multlined}
\end{aligned}
\end{equation}
where three $\rho_{i}$'s were parametrized in the second step as  
\begin{equation}
\rho_{1} = \cos(\theta_1)\sin(\theta_2)\,,\quad \rho_{2} = \sin(\theta_1) \sin(\theta_2)\,,\quad \rho_{3} = \cos(\theta_2)\,.
\end{equation}
A natural coframe is defined as
\begin{equation}
\begin{aligned}
\mathbf{e}^{\theta_{1}} &= \sin(\theta_2) \dd{\theta_{1}}\,, & \mathbf{e}^{\theta_{2}} &= \dd{\theta_{2}}\,,\\
\mathbf{e}^{\phi_{1}} &= \cos(\theta_1)\sin(\theta_2) \dd{\phi_{1}}\,, & \mathbf{e}^{\phi_{2}} &= \sin(\theta_1)\sin(\theta_2) \dd{\phi_{2}}\,,\quad \mathbf{e}^{\phi_{3}} = \cos(\theta_2) \dd{\phi_{3}}\,,
\end{aligned} 
\end{equation}
and thereby one finds the spin connections
\begin{equation}
\begin{aligned}
\omega^{\theta_{1}\theta_{2}}_{\theta_{1}} &= \cos(\theta_2)\,, & \omega^{\theta_{1}\phi_{1}}_{\phi_{1}} &= \sin(\theta_1)\,, & \omega^{\phi_{2}\theta_{1}}_{\phi_{2}} &= \cos(\theta_1)\,,\\
\omega^{\phi_{1}\theta_{2}}_{\phi_{1}} &= \cos(\theta_1)\cos(\theta_2)\,,& \omega^{\phi_{2}\theta_{2}}_{\phi_{2}} &= \sin(\theta_1)\cos(\theta_2)\,, & \omega^{\theta_{2}\phi_{3}}_{\phi_{3}} &= \sin(\theta_2)\,.
\end{aligned}
\end{equation}
Hence, the $S^{5}$ part of the gravitino equation has the following components:
\begin{equation}
\begin{aligned}
\partial_{\theta_{1}} \epsilon + \frac{1}{2}\cos(\theta_2)\Gamma_{\theta_{1}\theta_{2}}\epsilon &=\frac{i}{2}\sin(\theta_2)\Gamma_{\theta_{1}}\epsilon\,,\\
\partial_{\theta_{2}}\epsilon &= \frac{i}{2}\Gamma_{\theta_{2}}\epsilon\,,\\
\partial_{\phi_{1}}\epsilon + \frac{1}{2}\left(\sin(\theta_1)\Gamma_{\theta_{1}\phi_{1}}+ \cos(\theta_1)\cos(\theta_2)\Gamma_{\phi_{1}\theta_{2}}\right)\epsilon &= \frac{i}{2}\cos(\theta_1)\sin(\theta_2)\Gamma_{\phi_{1}}\epsilon\,,\\
\partial_{\phi_{2}}\epsilon + \frac{1}{2}\left(\cos(\theta_1)\Gamma_{\phi_{2}\theta_{1}}+\sin(\theta_1)\cos(\theta_2)\Gamma_{\phi_{2}\theta_{2}}\right)\epsilon &=\frac{i}{2}\sin(\theta_1)\sin(\theta_2)\Gamma_{\phi_{2}}\epsilon\,,\\
\partial_{\phi_{3}}\epsilon + \frac{1}{2}\sin(\theta_2)\Gamma_{\theta_{2}\phi_{3}}\epsilon &= \frac{i}{2}\cos(\theta_2)\Gamma_{\phi_{3}}\epsilon.
\end{aligned}
\end{equation}
Consequently, the Killing spinor on $S^{5}$ only is given by
\begin{equation}
\label{eq:spinor-s5}
\boxed{
\epsilon_{S^{5}} = e^{\frac{i}{2}\theta_{2}\Gamma_{\theta_{2}}}e^{\frac{i}{2}\phi_{3}\Gamma_{\phi_{3}}}e^{\frac{\theta_{1}}{2}\Gamma_{\theta_{2}\theta_{1}}}e^{\frac{\phi_{1}}{2}\Gamma_{\theta_{2}\phi_{1}}}e^{\frac{\phi_{2}}{2}\Gamma_{\theta_{1}\phi_{2}}}\tilde{\epsilon}_{0}}\,,
\end{equation}
where $\tilde{\epsilon}_{0}$ is {a constant spinor}, and all the Gamma matrices are flat. 
When combining the Killing spinors~\eqref{eq:spinor-ads5} and~\eqref{eq:spinor-s5} into the full solution on $\AdS{5}\times S^{5}$, one has to multiply the five-dimensional matrices $\Gamma_{\mu}$ and $\Gamma_{m}$ by $\sigma_{1}$ and by $\sigma_{2}$~\eqref{eq:gamma-decomposition}, respectively, so that we uplift them to the ten-dimensional matrices $\hat{\Gamma}_{A}$.
It is convenient to introduce the operator
\begin{equation}
\hat{\gamma} = -\sigma_{2} \otimes \mathbf{1}_{4\times4} \otimes \mathbf{1}_{4\times4}\,, \quad \hat{\gamma}^{2} = -\mathbf{1}_{32\times32}\,.
\end{equation} 
Consequently, the full form of the Killing spinor on the undeformed $\AdS{5} \times S^{5}$ is explicitly written as
\begin{multline}
\epsilon_{\AdS{5}\times S^{5}} = e^{\frac{i}{2}r\hat{\gamma}\Gamma_{r}}\left(1 + \frac{1}{2}x^{\alpha}\left( i \hat{\gamma}\hat{\Gamma}_{\alpha}+\hat{\Gamma}_{r}\hat{\Gamma}_{\alpha}\right)\right) \\
 \times e^{-\frac{1}{2}\theta_{2}\hat{\gamma}\hat{\Gamma}_{\theta_{2}}}e^{-\frac{1}{2}\phi_{3}\hat{\gamma}\hat{\Gamma}_{\phi_{3}}}e^{\frac{\theta_{1}}{2}\hat{\Gamma}_{\theta_{2}\theta_{1}}}e^{\frac{\phi_{1}}{2}\hat{\Gamma}_{\theta_{2}\phi_{1}}}e^{\frac{\phi_{2}}{2}\hat{\Gamma}_{\theta_{1}\phi_{2}}}\, \epsilon_{0}\,.
\end{multline}

\section{Killing spinors for undeformed $\AdS{7}\times S^{4}$}\label{app:KillingAdS7}

The $\AdS{7}\times S^{4}$ solution in eleven-dimensional supergravity is obtained by taking the four-form flux to be $F_{ijkl} = 3\epsilon_{ijkl}$, where $i,j,k,l$ range over the tangent space to the four-sphere. In the M-theory bulk, the Killing spinor obeys the following equation:
\begin{equation}
D_{A}\epsilon - \frac{1}{288}(\hat{\Gamma}_{APQRS}F^{PQRS} - 8 F_{APQR}\hat{\Gamma}^{PQR})\epsilon = 0\,.
\end{equation} 
As {in the case of} $\AdS{5} \times S^{5}$, we take the appropriate decomposition for the Gamma matrices~\eqref{eq:Gamma-Ads7} and the spinors~\eqref{eq:spinor-Ads7} and find two Killing spinor equations:
\begin{equation} 
\begin{aligned}
\AdS{7}:& \quad \ D_{\mu}\epsilon_{\AdS{7}} = \frac{1}{4}\Gamma_{\mu}\epsilon_{\AdS{7}}\\
S^{4}: \quad &\ D_{j}\epsilon_{S^{4}} = \frac{1}{2}\gamma\, \Gamma_{j}\epsilon_{S^{4}}\,,
\end{aligned}
\end{equation}
where the chirality operator is given by $\gamma = \Gamma_{89(10)(11)}$, and the Gamma matrices $\Gamma_{\mu}$ and $\Gamma_{j}$ are curved.

\subsection{$\AdS{7}$}
We start from the horospherical form of the metric,
\begin{equation}
\dd{s}^{2}_{\AdS{7}} = \dd{r}^{2} + e^{r}\eta_{\alpha\beta} \dd{x^{\alpha}} \dd{x^{\beta}}\,,
\end{equation}
where we note that the factor in the exponential is different from the one for $\AdS{5}$~\eqref{eq:AdS5HS}. This metric can be rewritten by introducing two polar coordinate systems:
\begin{equation}
\dd{s}^{2}_{\AdS{7}} = \dd{r}^{2} + e^{r} \pqty{-(\dd{x^{0}})^{2} + (\dd{x^{1}})^{2} + \dd{\rho_{1}}^{2} + \rho_{1}^{2} \dd{\varphi_{1}} + \dd{\rho_{2}}^{2} + \rho_{2}^{2} \dd{\varphi_{2}}^{2}}\,.
\end{equation}
In the above expression, the vielbein one-form is given by
\begin{equation}
\begin{aligned}
  \mathbf{e}^{0} &= e^{r/2} \dd{x^{0}}\,, &\mathbf{e}^{1} &= e^{r/2} \dd{x^{1}}\,,\\
  \mathbf{e}^{\rho_{i}} &= e^{r/2} \dd{\rho_{i}}\,, &\mathbf{e}^{\varphi_{i}} &=   e^{r/2}\rho_{i} \dd{\varphi_{i}}\,,\quad i=1,2\,.
\end{aligned}
\end{equation}
The non-zero components of the spin connection are
\begin{equation}
\begin{aligned}
\omega^{x^{\alpha}r}_{x^{\alpha}} &= \frac{1}{2}e^{r/2}\,,\quad  &\alpha &= 0,1\,,\quad i=1,2\,,\\
\omega^{\varphi_{i}\rho_{i}}_{\varphi_{i}} &= 1\,,\quad  &\omega^{\rho_{i}r}_{\rho_{i}} &= \frac{1}{2}e^{r/2}\,,\qquad \omega^{\varphi_{i}r}_{\varphi_{i}} = \frac{1}{2}e^{r/2}\rho_{i}\,.
\end{aligned}
\end{equation}
The corresponding Killing spinor equations become a bit more complicated:
\begin{equation}
\begin{aligned}
\partial_{x^{\alpha}}\epsilon  &= \frac{1}{4}e^{r/2}\Gamma_{x^{\alpha}}(1-\Gamma_{r})\epsilon\,,\quad \alpha=0,1\\  
\partial_{r}\epsilon &= \frac{1}{4}\Gamma_{r}\epsilon\,,\\
\partial_{\rho_{i}}\epsilon &= \frac{1}{4}e^{r/2}\Gamma_{\rho_{i}}(1-\Gamma_{r})\epsilon\,,\\
\partial_{\varphi_{i}}\epsilon &= \frac{1}{2}\Gamma_{\rho_{i}\varphi_{i}}\epsilon + \frac{1}{4}e^{r/2}\rho_{i}\Gamma_{\varphi_{i}}(1-\Gamma_{r})\epsilon\,,\qquad i=1,2\,,
\end{aligned}
\end{equation}
where all the Gamma matrices are flat. It is easy to read off the solution
\begin{equation}
\boxed{\epsilon_{\AdS{7}} = e^{\frac{1}{4}r\Gamma_{r}}\biggl[1+\frac{1}{4}\sum_{\substack{\alpha=0,1\\
                  i=1,2}}(x^{\alpha}\Gamma_{x^{\alpha}}+\rho_{i}\Gamma_{\rho_{i}})(1-\Gamma_{r})\biggr]e^{\frac{1}{2}\varphi_{1}\Gamma_{\rho_{1}\varphi_{1}}}e^{\frac{1}{2}\varphi_{2}\Gamma_{\rho_{2}\varphi_{2}}}\epsilon_{0}}\,,
\end{equation}
where $\epsilon_{0}$ is a constant 8-component spinor.

\subsection{$S^{4}$}

The metric on the four sphere looks quite similar to the five sphere up to the dependence on $\phi_{3}$. The geometry is parametrized as
\begin{equation}
\dd{s}^{2}_{S^{4}} = \dd{\theta_{2}}^{2} + \sin[2](\theta_2) \pqty{\dd{\theta_{1}}^{2} + \cos[2](\theta_{1}) \dd{\phi_{1}}^{2} + \sin[2](\theta_1) \dd{\phi_{2}}^{2}}\,,
\end{equation}
where there are manifestly two $U(1)$ isometry directions $\phi_{1}$ and $\phi_{2}$. The orthonormal basis can be naturally taken to be
\begin{equation}
\begin{aligned}
\mathbf{e}^{\theta_{1}} &= \sin(\theta_2)\,d\theta_{1}\,,\quad \mathbf{e}^{\theta_{2}} = d\theta_{2}\,,\\
\mathbf{e}^{\phi_{1}} &= \cos(\theta_1)\sin(\theta_2)\,d\phi_{1}\,,\quad \mathbf{e}^{\phi_{2}} = \sin(\theta_1)\sin(\theta_2)\,d\phi_{2}\,,
\end{aligned} 
\end{equation}
and the spin connections are found to be
\begin{equation}
\begin{aligned}
\omega^{\theta_{1}\theta_{2}}_{\theta_{1}} &= \cos(\theta_2)\,,\quad \omega^{\theta_{1}\phi_{1}}_{\phi_{1}} = \sin(\theta_1)\,,\quad \omega^{\phi_{2}\theta_{1}}_{\phi_{2}} = \cos(\theta_1)\,,\\
\omega^{\phi_{1}\theta_{2}}_{\phi_{1}} &= \cos(\theta_1)\cos(\theta_2)\,,\quad \omega^{\phi_{2}\theta_{2}}_{\phi_{2}} = \sin(\theta_1)\cos(\theta_2)\,,
\end{aligned}
\end{equation}
where $c_{i}$ and $s_{i}$ are understood in terms of $\theta_{i}$. Then the Killing spinor equations become
\begin{equation}
\begin{aligned}
\partial_{\theta_{1}}\epsilon + \frac{1}{2}\cos(\theta_2)\Gamma_{\theta_{1}\theta_{2}}\epsilon &= \frac{1}{2}\sin(\theta_2)\gamma\Gamma_{\theta_{1}}\epsilon\,,\\
\partial_{\theta_{2}}\epsilon &= \frac{1}{2} \gamma \Gamma_{\theta_{2}}\epsilon\,,\\
\partial_{\phi_{1}}\epsilon + \frac{1}{2}\sin(\theta_1)\Gamma_{\theta_{1}\phi_{1}}\epsilon + \frac{1}{2}\cos(\theta_1)\cos(\theta_2)\Gamma_{\phi_{1}\theta_{2}}\epsilon &= \frac{1}{2}\cos(\theta_1)\sin(\theta_2)\gamma\Gamma_{\phi_{1}}\epsilon\,,\\
\partial_{\phi_{2}}\epsilon + \frac{1}{2}\cos(\theta_1)\Gamma_{\phi_{2}\theta_{1}}\epsilon + \frac{1}{2}\sin(\theta_1)\cos(\theta_2)\Gamma_{\phi_{2}\theta_{2}}\epsilon &= \frac{1}{2}\sin(\theta_1)\sin(\theta_2)\gamma\Gamma_{\phi_{2}}\epsilon\,,
\end{aligned}
\end{equation}
where we used the flat Gamma matrices above. The solution takes the form
\begin{equation}
\boxed{
\epsilon_{S^{4}} = e^{\frac{1}{2}\theta_{2}\gamma \Gamma_{\theta_{2}}}e^{-\frac{1}{2}\theta_{1}\Gamma_{\theta_{1}\theta_{2}}}e^{\frac{1}{2}\phi_{1}\Gamma_{\theta_{2}\phi_{1}}}e^{\frac{1}{2}\phi_{2}\Gamma_{\theta_{1}\phi_{2}}}\epsilon_{0}}\,,
\end{equation}
where $\epsilon_{0}$ is a 4-component constant spinor.

Putting it all together, the Killing spinor on $\AdS{7}\times S^{4}$ is given by
\begin{multline}
\epsilon_{\AdS{7}\times S^{4}}
= e^{\frac{1}{4}r\hat{\gamma}\hat{\Gamma}_{r}} \biggl[ 1 + \frac{1}{4} \sum_{\substack{\alpha=0,1\\
    i=1,2}}\left(x^{\alpha}\left(  \hat{\gamma}\hat{\Gamma}_{\alpha}+\hat{\Gamma}_{r}\hat{\Gamma}_{\alpha}\right)+\rho_{i}\left(  \hat{\gamma}\hat{\Gamma}_{\rho_{i}}+\hat{\Gamma}_{r}\hat{\Gamma}_{\rho_{i}}\right)\right)\biggr] \\
\times e^{\frac{1}{2}\varphi_{1}\hat{\Gamma}_{\rho_{1}\varphi_{1}}} e^{\frac{1}{2}\varphi_{2}\hat{\Gamma}_{\rho_{2}\varphi_{2}}}
                 e^{\frac{1}{2}\theta_{2}\hat{\gamma} \hat{\Gamma}_{\theta_{2}}}e^{-\frac{1}{2}\theta_{1}\hat{\Gamma}_{\theta_{1}\theta_{2}}}e^{\frac{1}{2}\phi_{1}\hat{\Gamma}_{\theta_{2}\phi_{1}}}e^{\frac{1}{2}\phi_{2}\hat{\Gamma}_{\theta_{1}\phi_{2}}}\epsilon_{0}.
\end{multline}
\begin{small}
  \dosserif
  \bibliography{bibliography}{}
  \bibliographystyle{JHEP}
\end{small}

\end{document}